\documentclass[aps,pra,twocolumn,groupedaddress,nofootinbib,longbibliography]{revtex4-2}
\usepackage{graphicx} 

\usepackage{booktabs}
\usepackage{tikz}
\usepackage{pgfplots}
\usepackage{subcaption}
\usepackage{graphicx}
\usepackage{caption}
\usepackage{tabularx}
\usepackage{mathtools}
\usepackage{braket}
\usepackage[english]{babel}
\usepackage{amsthm}
\usepackage{algpseudocode}

\newcounter{algorithm}
\renewcommand{\thealgorithm}{\arabic{algorithm}}

\usepackage{xcolor, soul}
\usetikzlibrary{positioning, shapes.geometric, calc}
\usepackage{XCharter} 
\usepackage[T1]{fontenc}

\newcommand{\tr}{{\operatorname{tr}}}
\usepackage[charter,cal=cmcal,sfscaled=false]{mathdesign}
\definecolor{iffsred}{cmyk}{0.12,0.94,0.87,0.34}
\definecolor{uestcblue}{cmyk}{0.99,0.78,0.16,0.03}

\usepackage{hyperref}

\hypersetup{
  pdftitle={Symmetrized Block-Product Periodic Marginals in Infinite Translation-Invariant Quantum Chains},
  pdfauthor={Zeng, Yang, Zhang, Wang},
  pdfstartview=Fit,
  pdfpagelayout=SinglePage,
  colorlinks,
  linkcolor=uestcblue,
  citecolor=uestcblue,
  urlcolor=iffsred}

\begin{document}

\title{Symmetrized Block-Product Periodic Marginals in Infinite Translation-Invariant Quantum Chains}

\author{Xiao Zeng}
\affiliation{Institute of Fundamental and Frontier Sciences, University of Electronic Science and Technology of China, 611731, Chengdu, China}
\affiliation{Ministry of Education Key Laboratory of Quantum Physics and Photonic Quantum Information (University of Electronic Science and Technology of China), 611731, Chengdu, China}
\author{Kaiyan Yang}
\author{Lingxia Zhang}
\affiliation{Institute of Fundamental and Frontier Sciences, University of Electronic Science and Technology of China, 611731, Chengdu, China}
\affiliation{Ministry of Education Key Laboratory of Quantum Physics and Photonic Quantum Information (University of Electronic Science and Technology of China), 611731, Chengdu, China}
\author{Zizhu Wang}\email{zizhu@uestc.edu.cn}
\affiliation{Institute of Fundamental and Frontier Sciences, University of Electronic Science and Technology of China, 611731, Chengdu, China}
\affiliation{Ministry of Education Key Laboratory of Quantum Physics and Photonic Quantum Information (University of Electronic Science and Technology of China), 611731, Chengdu, China}

\begin{abstract}
We study local marginals in one-dimensional translation-invariant quantum systems that may hide finite-period structure. Given an $n$-site reduced density matrix, we ask whether it can be obtained by repeating a finite $p$-site block state along the chain and averaging over the $p$ lattice translations. This defines a symmetrized block-product periodic marginal problem, which provides a route both to diagnosing hidden periodic order from local data and to upper bounding ground-state energy densities of infinite translation-invariant local Hamiltonians. We develop two complementary methods. The first is a semidefinite-programming relaxation based on block permutation symmetry and positive partial transpose constraints, which outer-approximates the convex hull of such marginals and yields certified infeasibility tests. The second is a symmetrized matrix product state ansatz, which constructs explicit block-product periodic states and gives variational upper bounds. We benchmark the framework on the Majumdar–Ghosh model, transverse-field Ising, XX, XXZ, and contextuality-related spin models. The results show that the method captures the expected finite-period structure in exactly solvable cases and gives systematically improving variational energies as the period and bond dimension increase. We also formulate a periodic-NPA relaxation for translation-invariant contextuality witnesses and recover the known quantum limits in the tested examples.
\end{abstract}
\maketitle

\section{Introduction}

Translation invariance (TI) is one of the basic organizing principles of quantum many-body physics. In TI lattice systems, however, the practically relevant question is often not simply whether a state is one-site TI, but what kind of global state is compatible with the local data available to theory or experiment. This places the problem within the broader family of quantum marginal and $N$-representability problems, where one asks whether prescribed local reduced states are compatible with a global quantum state~\cite{klyachko2006quantum,liu2007quantum, kimEntropyScalingLaw2021,bravyi2004requirements}.

A key complication is that local marginals may hide finite-period order. Near-neighbor reduced states can be compatible not only with genuinely one-site TI states, but also with symmetry-broken phases whose fundamental period is larger and whose one-site TI is restored only after averaging over lattice translations. This distinction appears already in dimerized systems such as the Majumdar--Ghosh chain, whose spin-$1/2$ $J_1$--$J_2$ Heisenberg Hamiltonian has exactly dimerized ground states with a two-site unit cell at the Majumdar--Ghosh point~\cite{majumdarNextNearestNeighborInteractionLinear1969,majumdarNextNearestNeighborInteractionLinear1969a}. More broadly, translation-symmetry breaking is a standard mechanism in one-dimensional spin systems: competing-interaction Heisenberg chains exhibit well-characterized fluid--dimer transitions~\cite{Haldane1982,OKAMOTO1992433}, competing interactions can stabilize commensurate modulated phases with distinct periods, as in the axial next-nearest-neighbor Ising model and its devil's-staircase phenomenology~\cite{FisherSelke1980,Bak1982,Selke1988}, and Lieb--Schultz--Mattis-type commensurability constraints imply that certain gapped phases, such as magnetization plateaus, require an enlarged unit cell or ground-state degeneracy~\cite{LiebSchultzMattis1961,OshikawaYamanakaAffleck1997,Oshikawa2000,Hastings2004}. Related finite-period structures also arise in recent studies of entanglement, nonlocality, and contextuality in infinite one-dimensional TI systems~\cite{yangContextualityInfiniteOnedimensional2022,wangEntanglementNonlocalityInfinite2017}. More generally, the set of admissible reduced states of infinite TI chains is known to resist any exact finite semialgebraic characterization~\cite{blakajSetReducedStates2024}, making constructive approximation schemes essential.

Motivated by these observations, we study a symmetrized block-product periodic marginal problem: given a local reduced density matrix, determine whether it can arise from repeating a finite $p$-site block state along the infinite chain and then averaging over the $p$ lattice translations. This problem has two complementary motivations. The first is diagnostic. In quantum simulation and quantum-device experiments, full many-body tomography is generally infeasible, while local marginals are accessible. It is therefore natural to ask whether the observed local data are compatible with hidden finite-period order. The second motivation is variational. For an infinite local Hamiltonian, the ground-state energy density is determined by local reduced states subject to global compatibility constraints. Explicit block-product periodic ansatz can therefore provide variational upper bounds on TI ground-state energy densities, while convex relaxations of the same compatibility problem provide complementary lower bounds. 

Two methodological traditions make this problem approachable. On the one hand, semidefinite-programming relaxations provide systematically improvable outer approximations to difficult quantum compatibility constraints. Relevant examples include the Doherty--Parrilo--Spedalieri separability hierarchy~\cite{dohertyCompleteFamilySeparability2004}, the Navascués--Pironio--Acín hierarchy for quantum correlations~\cite{navascuesBoundingSetQuantum2007,navascuesConvergentHierarchySemidefinite2008}, and related device-independent moment-matrix constructions~\cite{moroderDeviceindependentEntanglementQuantification2013}. Closely related SDP approaches have also been developed for entanglement marginal problems, including translation-invariant one-dimensional systems~\cite{navascues2021entanglement}. On the other hand, tensor-network methods exploit the low-entanglement structure of one-dimensional ground states. The area-law perspective explains why matrix product states are efficient in the gapped setting~\cite{hastingsAreaLawOnedimensional2007, perez2007matrix}, and density-matrix renormalization group methods make this efficiency algorithmically practical~\cite{whiteDensitymatrixAlgorithmsQuantum1993,schollwock2011density}. These two toolkits suggest a natural division of labor: SDP methods for small but certified instances, and tensor-network variational methods for larger periods and bond dimensions.

This paper combines these perspectives into a framework for symmetrized block-product periodic extensions. The central new ingredient is the systematic imposition of a finite block-product periodic structure prior to translation averaging. This viewpoint leads to several distinct developments. First, it defines a local compatibility problem that is different from ordinary translation-invariant marginal compatibility. Second, it yields an SDP relaxation whose infeasibility certifies that a given local marginal cannot arise from a prescribed hidden block-product periodic explanation. Third, it gives a constructive symmetrized MPS ansatz that produces explicit periodic variational states. Finally, it motivates a periodic-NPA relaxation, which reproduces the extremal translation-invariant contextuality bounds of Ref.~\cite{yangContextualityInfiniteOnedimensional2022} in the tested examples. In this sense, periodic structure serves as an intermediate language connecting symmetry breaking, variational many-body methods, and device-independent correlation bounds.

The examples studied here show that this framework captures physically meaningful structure. For the Majumdar--Ghosh chain, whose ground-state manifold is built from dimerized patterns, the SDP hierarchy recovers the expected $2$-periodic extension and reproduces the exact ground-state energy density for compatible periods~\cite{majumdarNextNearestNeighborInteractionLinear1969,majumdarNextNearestNeighborInteractionLinear1969a}. For more general one-dimensional models, the symmetrized MPS method gives systematically improving variational behavior as the period and bond dimension are increased, while also revealing model-dependent convergence patterns. Finally, for TI contextuality witnesses of the types considered in Ref.~\cite{yangContextualityInfiniteOnedimensional2022}, the periodic-NPA construction reproduces the known quantum limits in the tested cases, suggesting that the block-product periodic ansatz can be sufficient to capture extremal TI correlations.

The rest of the paper is organized as follows. We first define symmetrized block-product periodic states and the associated marginal compatibility problem. We then develop an SDP hierarchy that gives outer relaxations of the convex hull of block-product periodic marginals, and present a symmetrized tensor-network method, focusing on MPS as the practical one-dimensional ansatz. After that, we test the framework on the Majumdar--Ghosh model and benchmark the symmetrized MPS method on representative local Hamiltonians. Finally, we apply the same block-product periodic viewpoint to TI contextuality witnesses through a periodic-NPA relaxation.

\section{The block-product periodic marginal problem} \label{sec:problem}

In an infinite chain, periodicity and block factorization are conceptually distinct. A general $p$-periodic state is one that is invariant under translation by $p$ sites. In this work, however, the constructions developed below are built for the narrower subclass obtained by repeating the same $p$-site density operator independently along the chain. To keep the problem statement aligned with the later SDP formulation, we therefore work with block-product $p$-periodic states. Concretely, let $\sigma_{[p]}$ be a density operator on $p$ contiguous sites, and write $$\Omega^{(p)} := \sigma_{[p]}^{\otimes\infty}$$ for the infinite chain obtained by repeating this unit cell on every block. We do not require $p$ to be the minimal period; if the exact period is a divisor of $p$, the state is still admissible in our formulation. The Majumdar–Ghosh model provides the guiding example: its symmetry-broken dimerized ground states have a two-site unit cell and are related by a one-site translation. 

\begin{figure}[htbp]
\centering
\includegraphics[width=0.9\linewidth]{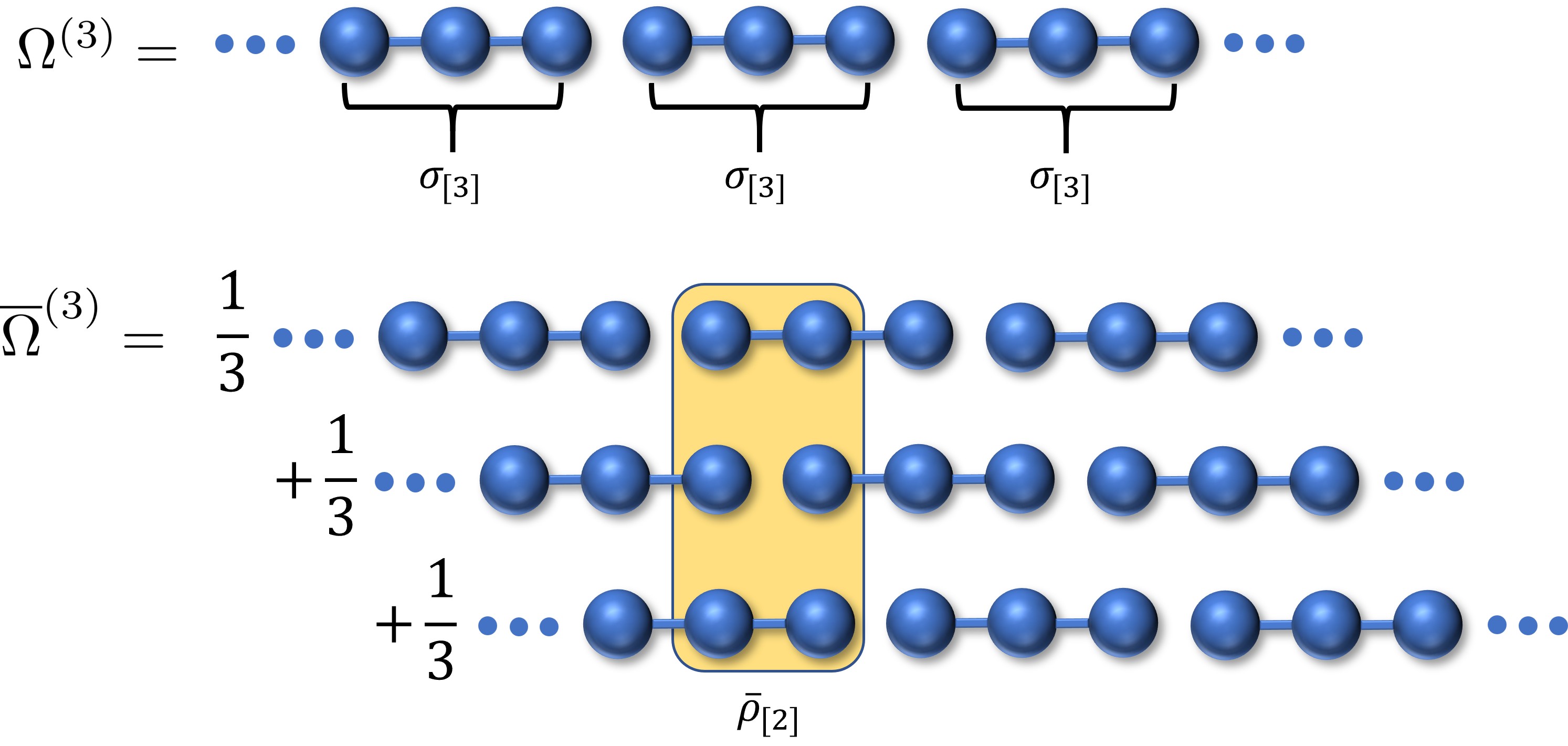}
\caption{An example of symmetrization when $p=3,r=2$. $\Omega^{(3)}$ is a periodic state with period $3$. The symmetrized state $\overline{\Omega}^{(3)}$ is one-site TI, and the yellow box shows the symmetrized state $\overline{\rho}_{[2]}$, which is composed by the reduced density matrices (RDMs) inside $\sigma_{[3]}$ and the RDM across the boundary.}
\label{fig:symmetrization1d}
\end{figure}

To associate with $\Omega^{(p)}$ a one-site TI description, we average over the translated copies at the level of states, \begin{equation}
\bar{\Omega}^{(p)} := \frac{1}{p}\sum_{s=0}^{p-1} T^s(\Omega^{(p)}),
\end{equation}
where $T$ is the one-site translation operator. For $r\le p$, the expression is particularly simple and is given by Eq.~\eqref{symmetrizing-state}. For $r>p$, the same definition is used by embedding the window into $k$ repeated blocks, with $kp\ge p+r-1$. The $r$-site reduced density matrix of the averaged state is \begin{equation}\label{symmetrizing-state}
    \bar{\rho}_{[r]}=\frac{1}{p}(\sum_{i=1}^{p-r+1}\sigma_{i:i+r-1}+\sum_{j=1}^{r-1}\sigma_{p-r+j+1:p}\otimes\sigma_{1:j}),
\end{equation} where $\sigma_{i:j}$ is the reduced density matrix of $\sigma_{[p]}$ on sites $i,\cdots,j$. Figure~\ref{fig:symmetrization1d} illustrates this construction for $p=3$ and $r=2$.

Given a reduced state $\rho_{[n]}$ on $n$ contiguous sites, the block-product periodic marginal problem asks whether there exist an integer $p$ and a $p$-site unit cell $\sigma_{[p]}$ such that $\rho_{[n]}=\bar{\rho}_{[n]},$ where $\bar{\rho}_{[n]}$ is obtained from the symmetrized infinite repetition of $\sigma_{[p]}$. Equivalently, the task is to decide whether the local data admit a block-product periodic extension whose one-site translation average matches the given state. This is a local-to-global compatibility problem for infinite TI systems. Recent work on admissible reduced states of infinite TI chains highlights both the importance of this viewpoint and the fact that exact finite descriptions of the relevant reduced-state sets are generally unavailable \cite{blakajSetReducedStates2024}. 

This problem is useful in at least two settings. First, it provides a diagnostic tool for local data. In quantum-simulation experiments, local reduced density matrices are often accessible, whereas full many-body tomography is infeasible. The block-product periodic marginal problem therefore gives a principled way to ask whether observed local data are compatible with hidden $p$-periodic order. Second, it provides a variational framework for infinite TI Hamiltonians. For a one-site TI nearest-neighbor Hamiltonian $H=\sum_i h_{i,i+1}$, the ground-state energy density is determined by admissible two-site reduced states. Explicit symmetrized block-product periodic ansatz therefore gives variational upper bounds on the exact energy density, and these bounds become tight when the ground-state manifold contains a compatible finite-period symmetry-broken state, as in the Majumdar--Ghosh example. In the following sections, we develop two complementary approaches: an SDP hierarchy that gives certified relaxations for small systems, and a symmetrized tensor-network ansatz that constructs explicit variational states for larger periods and bond dimensions.

\section{SDP approach to the block-product periodic marginal problem} \label{sec:sdp}

For a given local state $\rho_{[n]}$ defined on $n$ sites, the 1D block-product periodic marginal problem asks whether there exists a $p$-site state $\rho_{[p]}$ such that $\rho_{[n]}$ can be obtained as the symmetrized reduced density matrix of an infinite $p$-periodic block-product state generated from $\rho_{[p]}$.

A natural strategy is to fix a candidate period $p$ and search for such a state $\rho_{[p]}$. In principle, one can increase $p$ until a valid solution is found. However, a direct optimization over $\rho_{[p]}$ is not suitable for semidefinite programming (SDP), because the symmetrization map
$$
\rho_{[p]} \;\longmapsto\; \bar{\rho}_{[n]}
$$
is not linear. In particular, when the $n$-site window crosses the boundary of the $p$-site unit cell, the symmetrized reduced density matrix contains tensor-product contributions of reduced states of $\rho_{[p]}$, making the dependence nonlinear.

To bypass this difficulty, we lift the problem to a larger system.
For a given $p$, we introduce a state $\rho_{[kp]}$ defined on $kp$ sites. $k$ should satisfy the condition that $kp\geq p+n-1$. We divide the system into $k$ consecutive blocks, each consisting of $p$ sites. The key idea is that if a true $p$-periodic product structure exists, then the global state would take the form
$$
\rho_{[kp]} = \rho_{[p]}^{\otimes k}.
$$
This exact condition is nonconvex and cannot be imposed directly. Instead, we replace it with a convex relaxation.
We impose two conditions on $\rho_{[kp]}$:

1. \textbf{Block symmetry.}  
All $p$-site blocks are identical. This is enforced by requiring invariance under permutations of the $k$ blocks:
$$
\pi(\rho_{[kp]}) = \rho_{[kp]}, \quad \forall \pi \in S_k,
$$
where $S_k$ is the symmetric group acting on the $k$ blocks.

2. \textbf{Approximate separability between blocks.}  
In the exact product case, the blocks are fully separable. To approximate this condition, we use PPT-symmetric-extension constraints in the spirit of the DPS hierarchy~\cite{peres1996separability,horodecki2001separability,doherty2002distinguishing,dohertyCompleteFamilySeparability2004}. The DPS hierarchy requires that $\rho_{[kp]}$ remains positive under partial transposition with respect to every nontrivial subset of blocks:
$$
\rho_{[kp]}^{T_S} \succeq 0,
\quad
\forall \; \emptyset \neq S \subsetneq \{1,\dots,k\}.
$$
Increasing $k$ yields progressively tighter levels of the DPS hierarchy, which becomes complete in the limit $k\to\infty$ \cite{caves2002unknown, dohertyCompleteFamilySeparability2004}.
After lifting to $\rho_{[kp]}$, the symmetrization constraint becomes linear. The symmetrized $n$-site reduced density matrix can be written as
$$
\rho_{[n]} = \frac{1}{p} \sum_{i=1}^{p} \rho_{i,\dots,i+n-1},
$$
where $\rho_{i,\dots,i+n-1}$ are reduced density matrices of $\rho_{[kp]}$.

The block-product periodic marginal problem can now be relaxed to the following SDP feasibility problem:
\begin{equation}
  \begin{aligned}
\text{find} \quad & \rho_{[kp]} \\
\text{s.t.} \quad
& \rho_{[kp]} \succeq 0,\quad \tr(\rho_{[kp]}) = 1, \\
& \pi(\rho_{[kp]}) = \rho_{[kp]}, \quad \forall \pi \in S_k, \\
& \rho_{[kp]}^{T_S} \succeq 0, \quad \forall \emptyset \neq S \subsetneq \{1,\dots,k\}, \\
& \rho_{[n]} = \frac{1}{p} \sum_{i=1}^{p} \rho_{i,\dots,i+n-1}.
\end{aligned}
\end{equation}

This SDP provides an outer relaxation of the convex hull of symmetrized block-product periodic marginals. Consequently, if the SDP is \emph{infeasible}, then $\rho_{[n]}$ cannot admit a $p$-periodic block-product extension. If the SDP is \emph{feasible}, it only certifies compatibility with the chosen relaxation level: the solution $\rho_{[kp]}$ need not have the exact product form $\rho_{[p]}^{\otimes k}$, nor even correspond to a single underlying block state. In favorable cases, however, one may verify a posteriori whether the SDP solution factorizes, for example by checking whether $\lVert \rho_{[kp]}-\rho_{[p]}^{\otimes k} \rVert$
is below numerical tolerance in a suitable matrix norm. When such factorization is observed, the SDP value is realized by an explicit block-product state rather than only by the relaxation. The relaxation becomes tighter as $k$ increases, at the cost of rapidly growing computational complexity. In practice, small values of $k$, often $k=2$ or $3$, already provide useful tests and can be tight in benchmark examples.

For linear energy minimization, passing from the block-product set to its convex hull does not change the optimal value: a convex mixture cannot have lower energy than the lowest-energy component in the mixture. Thus the distinction between a single block-product state and a convex mixture is harmless at the level of the exact convexified variational optimum. The finite-level SDP, however, is larger than this convex hull and may therefore yield an energy below that attainable by any exact block-product state. Its optimal value should therefore be interpreted as a lower bound on the block-product variational energy, unless the optimizer is verified a posteriori to factorize, in which case the same value is achieved by an explicit block-product state.

\section{Symmetrized matrix product states}\label{sec:mps}
The semidefinite-programming approach is limited by the rapid growth of the Hilbert-space dimension, which restricts practical computations to relatively small systems. To improve scalability, we instead use matrix product states (MPS), which provide an efficient variational ansatz for one-dimensional quantum many-body states. Our construction is related in spirit to infinite-size MPS methods, where TI and finite unit cells are exploited directly in the thermodynamic limit~\cite{vidal2007classical}. Note that, unlike the SDP formulation, the MPS construction is restricted to block-product periodic states generated from pure p-site unit-cell states.

An MPS $\ket{\psi}$ on $N$ sites of local dimension $d$ takes the form
\begin{equation}
\ket{\psi}=\sum_{s_1,\cdots,s_N=1}^dA^{s_1}A^{s_2}\cdots A^{s_N}\ket{s_1s_2\cdots s_N},
\end{equation}
where $A^{s_k}$ is a matrix of dimension $(D_{k},D_{k+1})$, and $D_1=D_{N+1}=1$. The maximal dimension $D=\max_k D_k$ is called the bond dimension. A diagram representation is shown as
\begin{equation}
|\psi\rangle =
\begin{tikzpicture}[
    baseline=(A1.base),
    node distance=0.3cm,
    tensor/.style={
        draw,
        rectangle,
        rounded corners=4pt,
        minimum size=0.6cm,
        line width=0.8pt,
        font=\small,
        inner sep=2pt
    },
    wire/.style={
        line width=0.8pt,
        black
    },
    dot/.style={
        fill=black,
        circle,
        minimum size=2pt,
        inner sep=0pt
    }
]

    \node[tensor] (A1) at (0,0) {$A^{s_1}$};
    \node[tensor, right=of A1] (A2) {$A^{s_2}$};
    \node[tensor, right=of A2] (A3) {$A^{s_3}$};
    \node[tensor, right=of A3] (A4) {$A^{s_4}$};
    \node[tensor, right=1.2cm of A4] (AN) {$A^{s_N}$};

    \draw[wire] (A1) -- (A2);
    \draw[wire] (A2) -- (A3);
    \draw[wire] (A3) -- (A4);
    \draw[wire] (A4.east) -- ++(0.3,0);
    \draw[wire] (AN.west) -- ++(-0.3,0);

    \path (A4.east) -- (AN.west) coordinate[midway] (mid);
    \node[dot] at ($(mid)+(-0.2,0)$) {};
    \node[dot] at (mid) {};
    \node[dot] at ($(mid)+(0.2,0)$) {};

    \foreach \n in {A1,A2,A3,A4,AN} {
        \draw[wire] (\n.south) -- ++(0,-0.3);
    }

\end{tikzpicture}.
\end{equation}

The vertical leg of $A^{s_i}$ corresponds to the physical index, while the horizontal legs correspond to the virtual indices. The set of tensors $\{A^{s_i}\}$ uniquely determines the MPS wavefunction, but this representation is not unique: different tensor sets can represent the same quantum state. This redundancy is described by gauge transformations. For example, replacing $A^{s_2}$ by $X^{-1}A^{s_2}$ and $A^{s_1}$ by $A^{s_1}X$ leaves the physical state unchanged, because the two matrices $X$ and $X^{-1}$ cancel on the shared virtual bond. By choosing a suitable gauge, an MPS can be brought into left-orthonormal form, right-orthonormal form, or mixed-canonical form. The left-orthonormal form asks for each tensor $A^{s_i}$ ($i<N$) to satisfy 
\begin{equation}
   \sum_{s_i}(A^{s_i})^{\dagger}A^{s_i}=I, \text{ or } \begin{tikzpicture}[
 baseline={(current bounding box.center)},
 font=\small,
 tensor/.style={
   draw, rounded corners=2pt,
   minimum width=0.6cm, minimum height=0.6cm,
   inner sep=0pt,
   line width=0.8pt
 },
 edge/.style={line width=0.8pt}
]

\tikzset{
 parencurve/.style={edge, to path={to[out=180,in=180,looseness=1.0] (\tikztotarget)\tikztonodes}}
}
\def\ystep{1.0cm}
\def\rleg{0.3cm}

\node[tensor] (T) at (0,0) {$A^{s_i}$};
\node[tensor] (B) at (0,-\ystep) {$\bar A^{s_i}$};

\draw[edge] (T.south)--(B.north);
\draw[edge] (T.west) to[out=180,in=180,looseness=1.4] (B.west);

\draw[edge] (T.east)--++(\rleg,0);
\draw[edge] (B.east)--++(\rleg,0);

\node at (0.9cm,-0.5*\ystep) {$=$};

\begin{scope}[shift={(2.0cm,0)}] 
  \node[tensor, draw=none] (PT) at (0,0) {};
  \node[tensor, draw=none] (PB) at (0,-\ystep) {};

  \draw[edge] (PT.west) to[out=180,in=180,looseness=1.4] (PB.west);
\end{scope}
\end{tikzpicture}
\end{equation}
(there is no left arc for $A^{s_1}$).
The right-orthonormal form demands each tensor to satisfy 
\begin{equation}
   \sum_{s_i}A^{s_i}(A^{s_i})^{\dagger}=I, \text{ or } \begin{tikzpicture}[
 baseline={(current bounding box.center)},
 font=\small,
 tensor/.style={
   draw, rounded corners=2pt,
   minimum width=0.6cm, minimum height=0.6cm,
   inner sep=0pt,
   line width=0.8pt
 },
 edge/.style={line width=0.8pt}
]

\tikzset{
 parencurve/.style={edge, to path={to[out=180,in=180,looseness=1.0] (\tikztotarget)\tikztonodes}}
}
\def\ystep{1.0cm}
\def\rleg{0.3cm}
\def\lleg{-0.3cm}

\node[tensor] (T) at (0,0) {$A^{s_i}$};
\node[tensor] (B) at (0,-\ystep) {$\bar A^{s_i}$};

\draw[edge] (T.south)--(B.north);
\draw[edge] (T.east) to[out=0,in=0,looseness=1.4] (B.east);

\draw[edge] (T.west)--++(\lleg,0);
\draw[edge] (B.west)--++(\lleg,0);

\node at (1.0cm,-0.5*\ystep) {$=$};

\begin{scope}[shift={(1.7cm,0)}] 
  \node[tensor, draw=none] (PT) at (0,0) {};
  \node[tensor, draw=none] (PB) at (0,-\ystep) {};

  \draw[edge] (PT.west) to[out=0,in=0,looseness=1.4] (PB.west);
\end{scope}
\end{tikzpicture}.
\end{equation}
In the mixed-canonical form, we choose a center position $k$ and fix the gauge such that all tensors to its left are left-orthonormal, while all tensors to its right are right-orthonormal. We denote the left- and right-orthonormal tensors by $A_L^{s_i}$ and $A_R^{s_i}$, respectively.

The efficiency of an MPS representation is governed by the bond dimension $D$, which controls the amount of entanglement the ansatz can capture. Since ground states of one-dimensional local gapped Hamiltonians typically satisfy an area law for entanglement entropy \cite{hastingsAreaLawOnedimensional2007}, MPS provide a natural variational class for approximating such states. This is the basis of highly accurate algorithms such as the density matrix renormalization group \cite{whiteDensitymatrixAlgorithmsQuantum1993}.

To solve the block-product periodic marginal problem using MPS, we represent the $p$-site unit cell by an MPS $\ket{\psi}$. We then optimize $\ket{\psi}$ so that the symmetrized local reduced density matrix obtained from this unit cell agrees with the target state $\rho_{[n]}$. Since an $n$-site window may cross the boundary of the unit cell, it is convenient to work with two copies of the MPS. We therefore define $\ket{\phi}=\ket{\psi}\otimes\ket{\psi}$, whose diagrammatic representation is
\begin{equation}
    \ket{\phi} = 
    \begin{tikzpicture}[
        baseline=(A1.base),
        node distance=0.3cm,
        tensor/.style={
            draw,
            rectangle,
            rounded corners=4pt,
            minimum size=0.6cm,
            line width=0.8pt,
            font=\small,
            inner sep=2pt
        },
        wire/.style={
            line width=0.8pt,
            black
        },
        dot/.style={
            fill=black,
            circle,
            minimum size=2pt,
            inner sep=0pt
        }
    ]
    
        \node[tensor] (A1) at (0,0) {$A^{s_1}$};
        \node[tensor, right=of A1] (A2) {$A^{s_2}$};
        \node[tensor, right=1.2cm of A2] (Ap) {$A^{s_p}$};
        
        \node[tensor, right=0.3cm of Ap] (Ap1) {$A^{s_{p+1}}$};
        \node[tensor, right=of Ap1] (Ap2) {$A^{s_{p+2}}$};
        \node[tensor, right=1.2cm of Ap2] (A2p) {$A^{s_{2p}}$};

        \draw[wire] (A1) -- (A2);
        \draw[wire] (A2.east) -- ++(0.3,0);
        \draw[wire] (Ap.west) -- ++(-0.3,0);
        
        \draw[wire] (Ap1) -- (Ap2);
        \draw[wire] (Ap2.east) -- ++(0.3,0);
        \draw[wire] (A2p.west) -- ++(-0.3,0);

        \path (A2.east) -- (Ap.west) coordinate[midway] (mid1);
        \node[dot] at ($(mid1) - (0.2, 0)$) {};
        \node[dot] at (mid1) {};
        \node[dot] at ($(mid1) + (0.2, 0)$) {};

        \path (Ap2.east) -- (A2p.west) coordinate[midway] (mid2);
        \node[dot] at ($(mid2) - (0.2, 0)$) {};
        \node[dot] at (mid2) {};
        \node[dot] at ($(mid2) + (0.2, 0)$) {};

        \foreach \node in {A1, A2, Ap, Ap1, Ap2, A2p} {
            \draw[wire] (\node.south) -- ++(0,-0.3);
        }
    \end{tikzpicture}.
\end{equation}
The right-hand side of $A^{s_p}$ and the left-hand side of $A^{s_{p+1}}$ are not connected, indicating the MPS is separable across the middle cut.

The symmetrized $r$-site reduced density matrix is composed by the RDMs inside $\ket{\psi}$ and the RDMs across the middle cut. For example, for $p=4,r=2$, the symmetrized state (unnormalized) is shown as

\begin{widetext}
\begin{equation}
\begin{tikzpicture}[
  baseline={(current bounding box.center)},
  font=\small,
  scale=0.9,
  transform shape,
  tensor/.style={
    draw, rounded corners=2pt,
    minimum width=0.6cm, minimum height=0.6cm,
    inner sep=0pt,
    line width=0.8pt
  },
  edge/.style={line width=0.8pt},
  leg/.style={line width=0.8pt}
]
\def\xstep{0.8cm}   
\def\ystep{1.5cm}
\def\leglen{0.3cm}

\newcommand{\PlaceDoubleRow}{
  \node[tensor] (T1) at (0,0) {$A^{s_1}$};
  \node[tensor] (T2) at ($(T1)+(\xstep,0)$) {$A^{s_2}$};
  \node[tensor] (T3) at ($(T2)+(\xstep,0)$) {$A^{s_3}$};
  \node[tensor] (T4) at ($(T3)+(\xstep,0)$) {$A^{s_4}$};
  \node[tensor] (B1) at ($(T1)+(0,-\ystep)$) {$\bar A^{s_1}$};
  \node[tensor] (B2) at ($(T2)+(0,-\ystep)$) {$\bar A^{s_2}$};
  \node[tensor] (B3) at ($(T3)+(0,-\ystep)$) {$\bar A^{s_3}$};
  \node[tensor] (B4) at ($(T4)+(0,-\ystep)$) {$\bar A^{s_4}$};
  \draw[edge] (T1.east)--(T2.west) (T2.east)--(T3.west) (T3.east)--(T4.west);
  \draw[edge] (B1.east)--(B2.west) (B2.east)--(B3.west) (B3.east)--(B4.west);
}
\newcommand{\VConnect}[1]{\draw[edge] (T#1.south)--(B#1.north);}
\newcommand{\VDisconnect}[1]{%
  \draw[leg] (T#1.south)--++(0,-\leglen);
  \draw[leg] (B#1.north)--++(0,\leglen);
}

\node at (0cm,-0.7cm) {$\overline{\rho}_{[2]}=$};

\begin{scope}[shift={(0.8,0)}]
  \PlaceDoubleRow
  \VDisconnect{1}\VDisconnect{2}\VConnect{3}\VConnect{4}
\end{scope}

\node at (3.8cm,-0.7cm) {$+$};

\begin{scope}[shift={(4.3,0)}]
  \PlaceDoubleRow
  \VConnect{1}\VDisconnect{2}\VDisconnect{3}\VConnect{4}
\end{scope}

\node at (7.3cm,-0.7cm) {$+$};

\begin{scope}[shift={(7.8,0)}]
  \PlaceDoubleRow
  \VConnect{1}\VConnect{2}\VDisconnect{3}\VDisconnect{4}
\end{scope}

\node at (10.8cm,-0.7cm) {$+$};

\begin{scope}[shift={(11.3,0)}]
  \PlaceDoubleRow
  \VConnect{1}\VConnect{2}\VConnect{3}\VDisconnect{4}
  \draw[leg] (T4.south)--++(0,-\leglen);
\end{scope}


\begin{scope}[shift={(14.6,0)}]
  \PlaceDoubleRow
  \VDisconnect{1}\VConnect{2}\VConnect{3}\VConnect{4}
\end{scope}

\end{tikzpicture}
\end{equation}
\end{widetext}

The next step is to optimize the tensors $\{A^{s_i}\}$ by a sweep algorithm so that the symmetrized reduced density matrix generated from $\ket{\phi}$ best approximates the target state $\rho_{[r]}$. Equivalently, this fitting problem can be written as an energy minimization problem. For example, by choosing the local Hamiltonian term as $h=-\rho_{[2]}$, minimizing the energy density of $h$ over the symmetrized MPS ansatz favors states whose two-site reduced density matrix has large overlap with the target $\rho_{[2]}$. More generally, the same framework can be applied to physical local Hamiltonians, such as the Heisenberg or MG Hamiltonian. In this case, the variational energy obtained from the symmetrized MPS ansatz gives an upper bound on the true ground-state energy density of the infinite TI Hamiltonian. 

If the boundary-crossing term were absent, the optimization would reduce to the standard ground-state problem for a Hamiltonian with open boundary conditions (OBC), which can be efficiently treated by algorithms such as DMRG. The main additional difficulty in the present setting is the boundary term, which couples the end of the unit cell back to its beginning through the symmetrized reduced density matrix. The energy can be evaluated by contracting $h$ with $\overline{\rho}_{[2]}$, i.e., 
\begin{widetext}
\begin{equation} \label{eq:symmetrized energy}
\begin{tikzpicture}[
  baseline={(current bounding box.center)},
  font=\small,
  scale=0.9,
  transform shape,
  tensor/.style={
    draw, rounded corners=2pt,
    minimum width=0.6cm, minimum height=0.6cm,
    inner sep=0pt,
    line width=0.8pt
  },
  hbox/.style={
    draw, rounded corners=3pt,
    minimum width=1.5cm,
    minimum height=0.5cm,
    inner sep=0pt,
    line width=0.8pt
  },
  edge/.style={line width=0.8pt}
]
\def\xstep{0.8cm}
\def\ystep{1.5cm}
\def\leglen{0.3cm}

\newcommand{\PlaceDoubleRowFour}{
  \node[tensor] (T1) at (0,0) {$A^{s_1}$};
  \node[tensor] (T2) at ($(T1)+(\xstep,0)$) {$A^{s_2}$};
  \node[tensor] (T3) at ($(T2)+(\xstep,0)$) {$A^{s_3}$};
  \node[tensor] (T4) at ($(T3)+(\xstep,0)$) {$A^{s_4}$};

  \node[tensor] (B1) at ($(T1)+(0,-\ystep)$) {$\bar A^{s_1}$};
  \node[tensor] (B2) at ($(T2)+(0,-\ystep)$) {$\bar A^{s_2}$};
  \node[tensor] (B3) at ($(T3)+(0,-\ystep)$) {$\bar A^{s_3}$};
  \node[tensor] (B4) at ($(T4)+(0,-\ystep)$) {$\bar A^{s_4}$};

  \draw[edge] (T1.east)--(T2.west) (T2.east)--(T3.west) (T3.east)--(T4.west);
  \draw[edge] (B1.east)--(B2.west) (B2.east)--(B3.west) (B3.east)--(B4.west);
}

\newcommand{\PlaceDoubleRowFiveWrap}{
  \node[tensor] (T1) at (0,0) {$A^{s_1}$};
  \node[tensor] (T2) at ($(T1)+(\xstep,0)$) {$A^{s_2}$};
  \node[tensor] (T3) at ($(T2)+(\xstep,0)$) {$A^{s_3}$};
  \node[tensor] (T4) at ($(T3)+(\xstep,0)$) {$A^{s_4}$};
  \node[tensor] (T5) at ($(T4)+(\xstep,0)$) {$A^{s_1}$};
  \node[tensor] (T6) at ($(T5)+(\xstep,0)$) {$A^{s_2}$};
  \node[tensor] (T7) at ($(T6)+(\xstep,0)$) {$A^{s_3}$};
  \node[tensor] (T8) at ($(T7)+(\xstep,0)$) {$A^{s_4}$};

  \node[tensor] (B1) at ($(T1)+(0,-\ystep)$) {$\bar A^{s_1}$};
  \node[tensor] (B2) at ($(T2)+(0,-\ystep)$) {$\bar A^{s_2}$};
  \node[tensor] (B3) at ($(T3)+(0,-\ystep)$) {$\bar A^{s_3}$};
  \node[tensor] (B4) at ($(T4)+(0,-\ystep)$) {$\bar A^{s_4}$};
  \node[tensor] (B5) at ($(T5)+(0,-\ystep)$) {$\bar A^{s_1}$};
  \node[tensor] (B6) at ($(T6)+(0,-\ystep)$) {$\bar A^{s_2}$};
  \node[tensor] (B7) at ($(T7)+(0,-\ystep)$) {$\bar A^{s_3}$};
  \node[tensor] (B8) at ($(T8)+(0,-\ystep)$) {$\bar A^{s_4}$};

  \draw[edge] (T1.east)--(T2.west) (T2.east)--(T3.west) (T3.east)--(T4.west);
  \draw[edge] (T5.east)--(T6.west) (T6.east)--(T7.west) (T7.east)--(T8.west);

  \draw[edge] (B1.east)--(B2.west) (B2.east)--(B3.west) (B3.east)--(B4.west);
  \draw[edge] (B5.east)--(B6.west) (B6.east)--(B7.west) (B7.east)--(B8.west);
}

\newcommand{\VConnect}[1]{\draw[edge] (T#1.south)--(B#1.north);}

\newcommand{\InsertHBox}[2]{%
  \node[hbox] (H) at ($ (T#1)!0.5!(T#2) + (0,-0.5*\ystep) $) {$h$};

  \coordinate (Ht1) at ($(H.north west)!(T#1.south)!(H.north east)$);
  \coordinate (Ht2) at ($(H.north west)!(T#2.south)!(H.north east)$);
  \coordinate (Hb1) at ($(H.south west)!(B#1.north)!(H.south east)$);
  \coordinate (Hb2) at ($(H.south west)!(B#2.north)!(H.south east)$);

  \draw[edge] (T#1.south)--(Ht1);
  \draw[edge] (T#2.south)--(Ht2);
  \draw[edge] (Hb1)--(B#1.north);
  \draw[edge] (Hb2)--(B#2.north);
}

\node at (0cm,-0.7cm) {$e=$};

\begin{scope}[shift={(0.8,0)}]
  \PlaceDoubleRowFour
  \InsertHBox{1}{2}
  \VConnect{3}\VConnect{4}
\end{scope}

\node at (3.8cm,-0.7cm) {$+$};

\begin{scope}[shift={(4.3,0)}]
  \PlaceDoubleRowFour
  \VConnect{1}
  \InsertHBox{2}{3}
  \VConnect{4}
\end{scope}

\node at (7.3cm,-0.7cm) {$+$};

\begin{scope}[shift={(7.8,0)}]
  \PlaceDoubleRowFour
  \VConnect{1}\VConnect{2}
  \InsertHBox{3}{4}
\end{scope}

\node at (10.8cm,-0.7cm) {$+$};

\begin{scope}[shift={(11.3,0)}]
  \PlaceDoubleRowFiveWrap
  \VConnect{1}\VConnect{2}\VConnect{3}\VConnect{6}\VConnect{7}\VConnect{8}
  \InsertHBox{4}{5}
\end{scope}

\node at (17.8cm,-0.7cm) {.};

\end{tikzpicture}
\end{equation}
\end{widetext}

The tensors $\{A^{s_i}\}$ are optimized sequentially using a sweep algorithm. At each step, we fix all tensors except one and update the remaining tensor. Without loss of generality, we describe the update of $A^{s_1}$ and assume that the MPS is in mixed form centered at $A^{s_1}$; the other tensors are updated analogously. The derivative of $e$ with respect to $\bar{A}^{s_1}$ is 
\begin{widetext}
\begin{equation}\label{eq:derivative}
\begin{tikzpicture}[
  baseline={(current bounding box.center)},
  font=\small,
  scale=0.9,
  transform shape,
  tensor/.style={
    draw, rounded corners=2pt,
    minimum width=0.6cm, minimum height=0.6cm,
    inner sep=0pt,
    line width=0.8pt
  },
  blank/.style={
    draw=none, rounded corners=0pt,
    minimum width=0.6cm, minimum height=0.6cm,
    inner sep=0pt,
    line width=0pt
  },
  hbox/.style={
    draw, rounded corners=3pt,
    minimum width=1.5cm,
    minimum height=0.5cm,
    inner sep=0pt,
    line width=0.8pt
  },
  edge/.style={line width=0.8pt},
]
\def\xstep{0.8cm}
\def\ystep{1.5cm}
\def\rowgap{2.7cm}

\newcommand{\PlaceDoubleRowFour}{
  \node[tensor] (T1) at (0,0) {$A^{s_1}$};
  \node[tensor] (T2) at ($(T1)+(\xstep,0)$) {$A^{s_2}$};
  \node[tensor] (T3) at ($(T2)+(\xstep,0)$) {$A^{s_3}$};
  \node[tensor] (T4) at ($(T3)+(\xstep,0)$) {$A^{s_4}$};

  \node[tensor] (B1) at ($(T1)+(0,-\ystep)$) {$\bar A^{s_1}$};
  \node[tensor] (B2) at ($(T2)+(0,-\ystep)$) {$\bar A^{s_2}$};
  \node[tensor] (B3) at ($(T3)+(0,-\ystep)$) {$\bar A^{s_3}$};
  \node[tensor] (B4) at ($(T4)+(0,-\ystep)$) {$\bar A^{s_4}$};

  \draw[edge] (T1.east)--(T2.west) (T2.east)--(T3.west) (T3.east)--(T4.west);
  \draw[edge] (B1.east)--(B2.west) (B2.east)--(B3.west) (B3.east)--(B4.west);
}

\newcommand{\PlaceDoubleRowFourDeleteBone}{
  \node[tensor] (T1) at (0,0) {$A^{s_1}$};
  \node[tensor] (T2) at ($(T1)+(\xstep,0)$) {$A^{s_2}$};
  \node[tensor] (T3) at ($(T2)+(\xstep,0)$) {$A^{s_3}$};
  \node[tensor] (T4) at ($(T3)+(\xstep,0)$) {$A^{s_4}$};

  \node[blank]  (B1) at ($(T1)+(0,-\ystep)$) {};
  \node[tensor] (B2) at ($(T2)+(0,-\ystep)$) {$\bar A^{s_2}$};
  \node[tensor] (B3) at ($(T3)+(0,-\ystep)$) {$\bar A^{s_3}$};
  \node[tensor] (B4) at ($(T4)+(0,-\ystep)$) {$\bar A^{s_4}$};

  \draw[edge] (T1.east)--(T2.west) (T2.east)--(T3.west) (T3.east)--(T4.west);
  \draw[edge] (B2.east)--(B3.west) (B3.east)--(B4.west);
  \draw[edge] (B2.west)--++(-0.3,0);
}

\newcommand{\PlaceDoubleRowFiveWrap}{
  \node[tensor] (T1) at (0,0) {$A^{s_1}$};
  \node[tensor] (T2) at ($(T1)+(\xstep,0)$) {$A^{s_2}$};
  \node[tensor] (T3) at ($(T2)+(\xstep,0)$) {$A^{s_3}$};
  \node[tensor] (T4) at ($(T3)+(\xstep,0)$) {$A^{s_4}$};
  \node[tensor] (T5) at ($(T4)+(\xstep,0)$) {$A^{s_1}$};
  \node[tensor] (T6) at ($(T5)+(\xstep,0)$) {$A^{s_2}$};
  \node[tensor] (T7) at ($(T6)+(\xstep,0)$) {$A^{s_3}$};
  \node[tensor] (T8) at ($(T7)+(\xstep,0)$) {$A^{s_4}$};

  \node[tensor] (B1) at ($(T1)+(0,-\ystep)$) {$\bar A^{s_1}$};
  \node[tensor] (B2) at ($(T2)+(0,-\ystep)$) {$\bar A^{s_2}$};
  \node[tensor] (B3) at ($(T3)+(0,-\ystep)$) {$\bar A^{s_3}$};
  \node[tensor] (B4) at ($(T4)+(0,-\ystep)$) {$\bar A^{s_4}$};
  \node[tensor] (B5) at ($(T5)+(0,-\ystep)$) {$\bar A^{s_1}$};
  \node[tensor] (B6) at ($(T6)+(0,-\ystep)$) {$\bar A^{s_2}$};
  \node[tensor] (B7) at ($(T7)+(0,-\ystep)$) {$\bar A^{s_3}$};
  \node[tensor] (B8) at ($(T8)+(0,-\ystep)$) {$\bar A^{s_4}$};

  \draw[edge] (T1.east)--(T2.west) (T2.east)--(T3.west) (T3.east)--(T4.west);
  \draw[edge] (T5.east)--(T6.west) (T6.east)--(T7.west) (T7.east)--(T8.west);
  \draw[edge] (B1.east)--(B2.west) (B2.east)--(B3.west) (B3.east)--(B4.west);
  \draw[edge] (B5.east)--(B6.west) (B6.east)--(B7.west) (B7.east)--(B8.west);
}

\newcommand{\PlaceDoubleRowFiveWrapDeleteBoneleft}{
  \node[tensor] (T1) at (0,0) {$A^{s_1}$};
  \node[tensor] (T2) at ($(T1)+(\xstep,0)$) {$A^{s_2}$};
  \node[tensor] (T3) at ($(T2)+(\xstep,0)$) {$A^{s_3}$};
  \node[tensor] (T4) at ($(T3)+(\xstep,0)$) {$A^{s_4}$};
  \node[tensor] (T5) at ($(T4)+(\xstep,0)$) {$A^{s_1}$};
  \node[tensor] (T6) at ($(T5)+(\xstep,0)$) {$A^{s_2}$};
  \node[tensor] (T7) at ($(T6)+(\xstep,0)$) {$A^{s_3}$};
  \node[tensor] (T8) at ($(T7)+(\xstep,0)$) {$A^{s_4}$};

  \node[blank]  (B1) at ($(T1)+(0,-\ystep)$) {};
  \node[tensor] (B2) at ($(T2)+(0,-\ystep)$) {$\bar A^{s_2}$};
  \node[tensor] (B3) at ($(T3)+(0,-\ystep)$) {$\bar A^{s_3}$};
  \node[tensor] (B4) at ($(T4)+(0,-\ystep)$) {$\bar A^{s_4}$};
  \node[tensor] (B5) at ($(T5)+(0,-\ystep)$) {$\bar A^{s_1}$};
  \node[tensor] (B6) at ($(T6)+(0,-\ystep)$) {$\bar A^{s_2}$};
  \node[tensor] (B7) at ($(T7)+(0,-\ystep)$) {$\bar A^{s_3}$};
  \node[tensor] (B8) at ($(T8)+(0,-\ystep)$) {$\bar A^{s_4}$};

  \draw[edge] (T1.east)--(T2.west) (T2.east)--(T3.west) (T3.east)--(T4.west);
  \draw[edge] (T5.east)--(T6.west) (T6.east)--(T7.west) (T7.east)--(T8.west);
  \draw[edge] (B1.east)--(B2.west) (B2.east)--(B3.west) (B3.east)--(B4.west);
  \draw[edge] (B5.east)--(B6.west) (B6.east)--(B7.west) (B7.east)--(B8.west);
}

\newcommand{\PlaceDoubleRowFiveWrapDeleteBoneright}{
  \node[tensor] (T1) at (0,0) {$A^{s_1}$};
  \node[tensor] (T2) at ($(T1)+(\xstep,0)$) {$A^{s_2}$};
  \node[tensor] (T3) at ($(T2)+(\xstep,0)$) {$A^{s_3}$};
  \node[tensor] (T4) at ($(T3)+(\xstep,0)$) {$A^{s_4}$};
  \node[tensor] (T5) at ($(T4)+(\xstep,0)$) {$A^{s_1}$};
  \node[tensor] (T6) at ($(T5)+(\xstep,0)$) {$A^{s_2}$};
  \node[tensor] (T7) at ($(T6)+(\xstep,0)$) {$A^{s_3}$};
  \node[tensor] (T8) at ($(T7)+(\xstep,0)$) {$A^{s_4}$};

  \node[tensor] (B1) at ($(T1)+(0,-\ystep)$) {$\bar A^{s_1}$};
  \node[tensor] (B2) at ($(T2)+(0,-\ystep)$) {$\bar A^{s_2}$};
  \node[tensor] (B3) at ($(T3)+(0,-\ystep)$) {$\bar A^{s_3}$};
  \node[tensor] (B4) at ($(T4)+(0,-\ystep)$) {$\bar A^{s_4}$};
  \node[blank]  (B5) at ($(T5)+(0,-\ystep)$) {};
  \node[tensor] (B6) at ($(T6)+(0,-\ystep)$) {$\bar A^{s_2}$};
  \node[tensor] (B7) at ($(T7)+(0,-\ystep)$) {$\bar A^{s_3}$};
  \node[tensor] (B8) at ($(T8)+(0,-\ystep)$) {$\bar A^{s_4}$};

  \draw[edge] (T1.east)--(T2.west) (T2.east)--(T3.west) (T3.east)--(T4.west);
  \draw[edge] (T5.east)--(T6.west) (T6.east)--(T7.west) (T7.east)--(T8.west);
  \draw[edge] (B1.east)--(B2.west) (B2.east)--(B3.west) (B3.east)--(B4.west);
  \draw[edge] (B5.east)--(B6.west) (B6.east)--(B7.west) (B7.east)--(B8.west);
}

\newcommand{\VConnect}[1]{\draw[edge] (T#1.south)--(B#1.north);}

\newcommand{\InsertHBox}[2]{%
  \node[hbox] (H) at ($ (T#1)!0.5!(T#2) + (0,-0.5*\ystep) $) {$h$};

  \coordinate (Ht1) at ($(H.north west)!(T#1.south)!(H.north east)$);
  \coordinate (Ht2) at ($(H.north west)!(T#2.south)!(H.north east)$);
  \coordinate (Hb1) at ($(H.south west)!(B#1.north)!(H.south east)$);
  \coordinate (Hb2) at ($(H.south west)!(B#2.north)!(H.south east)$);

  \draw[edge] (T#1.south)--(Ht1);
  \draw[edge] (T#2.south)--(Ht2);
  \draw[edge] (Hb1)--(B#1.north);
  \draw[edge] (Hb2)--(B#2.north);
}

\newcommand{\InsertHBoxDeleteBone}[2]{%
  \node[hbox] (H) at ($ (T#1)!0.5!(T#2) + (0,-0.5*\ystep) $) {$h$};

  \coordinate (Ht1) at ($(H.north west)!(T#1.south)!(H.north east)$);
  \coordinate (Ht2) at ($(H.north west)!(T#2.south)!(H.north east)$);
  \coordinate (Hb1) at ($(H.south west)!(B#1.north)!(H.south east)$);
  \coordinate (Hb2) at ($(H.south west)!(B#2.north)!(H.south east)$);

  \draw[edge] (T#1.south)--(Ht1);
  \draw[edge] (T#2.south)--(Ht2);
  \draw[edge] (Hb1)--(B#1.north);
  \draw[edge] (Hb2)--(B#2.north);
}

\node at (0cm,-0.7cm) {{\large $\frac{\partial e}{\partial \bar{A}^{s_{1}}}$}$=$};

\begin{scope}[shift={(0.8cm,0)}]
  \PlaceDoubleRowFourDeleteBone
  \InsertHBoxDeleteBone{1}{2}
  \VConnect{3}\VConnect{4}
\end{scope}

\node at (3.8cm,-0.7cm) {$+$};

\begin{scope}[shift={(4.3cm,0)}]
  \PlaceDoubleRowFourDeleteBone
  \VConnect{1}
  \InsertHBox{2}{3}
  \VConnect{4}
\end{scope}

\node at (7.3cm,-0.7cm) {$+$};

\begin{scope}[shift={(7.8cm,0)}]
  \PlaceDoubleRowFourDeleteBone
  \VConnect{1}\VConnect{2}
  \InsertHBox{3}{4}
\end{scope}

\node at (10.8cm,-0.7cm) {$+$};

\begin{scope}[shift={(11.3cm,0)}]
  \PlaceDoubleRowFiveWrapDeleteBoneleft
  \VConnect{1}\VConnect{2}\VConnect{3}\VConnect{6}\VConnect{7}\VConnect{8}
  \InsertHBox{4}{5}
\end{scope}

\node at (0.1cm,-\rowgap-0.8cm) {$+$};

\begin{scope}[shift={(0.7cm,-\rowgap)}]
  \PlaceDoubleRowFiveWrapDeleteBoneright
  \VConnect{1}\VConnect{2}\VConnect{3}\VConnect{6}\VConnect{7}\VConnect{8}
  \InsertHBox{4}{5}
\end{scope}
\node at (7.7cm,-\rowgap-0.8cm) {.};
\end{tikzpicture}
\end{equation}
\end{widetext}

Eq.~\eqref{eq:derivative} defines the effective Hamiltonian $H_{A,1}$ such that 
\begin{equation} \label{eq:effective ha}
H_{\text{eff}}^{(1)}(A^{s_1})=\frac{\partial e}{\partial \bar{A}^{s_{1}}}.
\end{equation} 
The effective normalization matrix \(N_{\mathrm{eff}}^{(1)}\) is obtained from the
corresponding norm network, or equivalently by replacing the local
Hamiltonian insertion in the local energy network by the identity while
leaving the same local ket and bra indices open. Thus
\(H_{\mathrm{eff}}^{(1)}\) and \(N_{\mathrm{eff}}^{(1)}\) are linear maps on
the local tensor space of \(A_1\), and the one-site variational problem has
the generalized eigenvalue form
\[
H_{\mathrm{eff}}^{(1)}(A_1)
=
\lambda N_{\mathrm{eff}}^{(1)}(A_1).
\]
Since the MPS is brought into mixed-canonical form with \(A^{s_1}\) as the
orthogonality center before the update, the left and right norm environments
are identities. Therefore \(N_{\mathrm{eff}}^{(1)}\) reduces to the identity
map, and the update becomes
\[
H_{\mathrm{eff}}^{(1)}[A_1]=\lambda A_1 .
\]
We then update \(A^{s_1}\) by choosing the eigen-tensor associated with the
smallest eigenvalue. The same procedure is applied sequentially to the
remaining tensors after moving the orthogonality center. The sweep procedure is summarized in Algorithm~\ref{alg:sym-mps}.
\begin{center}
\begin{minipage}{1\linewidth}
\refstepcounter{algorithm}
\textbf{Algorithm~\thealgorithm. Symmetrized MPS optimization in the mixed-canonical form}
\label{alg:sym-mps}
\vspace{0.5em}
\begin{algorithmic}[1]
\State \textbf{Input:} local Hamiltonian term $h$ acting on $r$ sites, trial period $p\ge r$, bond dimension $D$, maximum number of sweeps $N_{\rm sweep}$, tolerance $\epsilon$
\State \textbf{Output:} optimized tensors $\{A_i\}_{i=1}^p$ and symmetrized energy density $e_p(D)$

\State Initialize a random $p$-site open-boundary MPS $\ket{\psi(A)}$. Bring $\ket{\psi(A)}$ into mixed-canonical form.
\State Compute the symmetrized energy density according to Eq.~\eqref{eq:symmetrized energy}.

\For{$s=1,\ldots,N_{\rm sweep}$}
    \State Set $e_{\rm old}\leftarrow e(A)$.

    \For{$i=1,\ldots,p$}
        \State Move the orthogonality center to site $i$.
        \State Construct the effective Hamiltonian $H_{\text{eff}}^{(i)}$ (Eq.~\eqref{eq:effective ha}) for tensor $A_i$.
        \State Update $A_i$ by choosing the eigenvector of $H_{\text{eff}}^{(i)}$ with the smallest eigenvalue.
        \State Restore the canonical form and move the orthogonality center to the next site.
    \EndFor

    \For{$i=p,\ldots,1$}
        \State Move the orthogonality center to site $i$.
        \State Construct the effective Hamiltonian $H_{\text{eff}}^{(i)}$ for tensor $A_i$.
        \State Update $A_i$ by choosing the eigenvector of $H_{\text{eff}}^{(i)}$ with the smallest eigenvalue.
        \State Restore the canonical form and move the orthogonality center to the previous site.
    \EndFor

    \State Compute $e_{\rm new}\leftarrow e(A)$.

    \If{$|e_{\rm new}-e_{\rm old}|<\epsilon$}
        \State \textbf{break}
    \EndIf
\EndFor

\State Set $e_p(D)\leftarrow e(A)$.
\State \textbf{return} optimized tensors $\{A_i\}_{i=1}^p$ and energy density $e_p(D)$.
\end{algorithmic}
\end{minipage}
\end{center}

After sufficiently many sweeps, the energy density typically converges to a stable variational minimum. The resulting optimized MPS therefore provides a candidate $p$-periodic state whose symmetrized reduced density matrix approximates the target local state, or equivalently a variational upper bound on the ground-state energy density of the corresponding TI Hamiltonian. Increasing $D$ enlarges the MPS variational manifold, allowing each $p$-site unit cell to capture more entanglement and represent more general local structures. Increasing $p$ plays a complementary role by permitting longer unit cells, which can accommodate finite-period patterns and longer-range correlation structures that are inaccessible to smaller blocks. Although the variational classes associated with arbitrary values of $p$ are not strictly nested, smaller-period solutions can be embedded into larger unit cells when $p$ is increased along appropriate multiples. Therefore, by taking both $p$ and $D$ sufficiently large, the symmetrized MPS ansatz is expected to provide increasingly tight variational upper bounds on the ground-state energy density, converging to the exact value in the limit $p\to\infty$. In the next section, we test this construction on representative one-dimensional models.

\section{Test cases and applications}\label{sec:app}
\subsection{A test case}
To test the SDP approach, we use the Majumdar--Ghosh (MG) model as a benchmark \cite{majumdarNextNearestNeighborInteractionLinear1969, majumdarNextNearestNeighborInteractionLinear1969a}. The MG model is a one-dimensional spin-$1/2$ chain with nearest-neighbor and next-nearest-neighbor interactions. Its Hamiltonian is
\begin{equation}
    H_{\text{MG}}=\sum_i \vec S_i \cdot \vec S_{i+1} + \frac{1}{2}\vec S_i \cdot \vec S_{i+2},
\end{equation}
where $\vec S_i$ denotes the spin-$1/2$ operator on site $i$. This model is particularly suitable for our purpose because its ground-state structure is exactly known: the ground space is spanned by two dimerized states, each formed by singlets on alternating nearest-neighbor bonds. These two states are related by a one-site translation and have period $2$.

Formally, after symmetrizing over the two translated dimer patterns, the two-site reduced density matrix is
\begin{equation}
  \rho_{[2]}=\left (\begin{array}{rrrr}
    \frac{1}{8} & 0 & 0 & 0\\
    0 & \frac{3}{8} & -\frac{1}{4} & 0\\
    0 & -\frac{1}{4} & \frac{3}{8} & 0\\
    0 & 0 & 0 & \frac{1}{8}
    \end{array} \right ).
\end{equation}
This provides a test instance for the block-product periodic marginal problem: the local state $\rho_{[2]}$ is expected to admit a $2$-periodic extension. We can verify this by checking the feasibility of the following SDP: \begin{equation}\label{sdp:mg}
\begin{aligned}\text{find} \quad &\rho_{[4]}\\
\text{subject to} \quad &\rho_{[4]} \succeq 0, \tr(\rho_{[4]})=1,\\
&\frac{1}{2}(\tr_{3,4}(\rho_{[4]}) + \tr_{1,4}(\rho_{[4]})) = \rho_{[2]},\\
&\rho_{[4]}^{T_L} \succeq 0,S(\rho_{[4]}) = \rho_{[4]},
\end{aligned}
\end{equation}
where $\rho_{[4]}^{T_L}$ is the PPT of $\rho_{[4]}$ with respect to the first two sites, and $S$ is the swap operator between the first two sites and the last two sites. Strictly speaking, the SDP hierarchy requires infinitely many copies to certify separability, and hence to rigorously establish the existence of a periodic extension. In the present example, however, once a feasible $\rho_{[4]}$ is obtained, we can directly check whether it factorizes into two identical two-site blocks. This provides an explicit verification of the desired $2$-periodic extension.

We solve Eq.~\eqref{sdp:mg} using MOSEK \cite{mosek} with a numerical tolerance of $10^{-8}$. The resulting feasible state $\rho_{[4]}$ satisfies
\begin{equation}
\rho_{[4]} = \tr_{3,4}(\rho_{[4]}) \otimes \tr_{1,2}(\rho_{[4]}),\quad
\tr_{3,4}(\rho_{[4]}) = \tr_{1,2}(\rho_{[4]}),
\end{equation}
up to an error below $10^{-8}$ in Frobenius norm.
In fact, the analytical solution is 
\begin{equation}  
  \tr_{3,4}(\rho_{[4]}) = \tr_{1,2}(\rho_{[4]}) =
\left(
\begin{array}{rrrr}
    0 & 0 & 0 & 0\\
    0 & \frac{1}{2} & -\frac{1}{2} & 0\\
    0 & -\frac{1}{2} & \frac{1}{2} & 0\\
    0 & 0 & 0 & 0
    \end{array}\right).
\end{equation} This explicitly verifies that $\rho_{[2]}$ admits a $2$-periodic extension.

We can also use the same SDP framework variationally, without assuming prior knowledge of the MG ground state. Instead of fixing the target two-site state $\rho_{[2]}$, we minimize the expectation value of the local MG Hamiltonian over the relaxed set of symmetrized $p$-periodic block-product states. This gives the following SDP: \begin{equation}\label{sdp:mg_energy}
\begin{aligned}\min_{\rho_{[2p]}} \quad &\tr(h \bar{\rho}_3)\\
\text{subject to} \quad &\rho_{[2p]} \succeq 0, \quad \tr(\rho_{[2p]})=1,\\
&\rho_{[2p]}^{T_L} \succeq 0,\quad S(\rho_{[2p]}) = \rho_{[2p]}, \\
& \frac{1}{p}\sum_{i=1}^p \rho_{i:i+3-1} = \bar{\rho}_3,
\end{aligned}
\end{equation}
where $h = \vec S_i \cdot \vec S_{i+1} + \frac{1}{2}\vec S_i \cdot \vec S_{i+2}$ is the local Hamiltonian term, $\rho_{[2p]}^{T_L}$ is the PPT of $\rho_{[2p]}$ with respect to the first $p$ sites, and $S$ is the swap operator between the first $p$ sites and the last $p$ sites. Table~\ref{tab:mg_energy} reports the optimal values of Eq.~\eqref{sdp:mg_energy} for different periods $p$. For even periods $p=2,4,6$, the SDP returns the exact value $-0.375$ and the states $\rho_{[2p]}$ factorize as block-product states, reflecting the $2$-periodic dimerized structure of the MG ground space.

\begin{table}[htbp]
  \caption{The optimal values of Eq.~\eqref{sdp:mg_energy} for different values of $p$. The exact ground state energy density of the MG model is $-0.375$.}
  \label{tab:mg_energy}
\centering
\begin{tabular}{|c|c|c|c|c|c|}
\hline
$p$ & 2 & 3 & 4 & 5 & 6 \\
\hline
$e_p$ & -0.375 & -0.310537 & -0.375 & -0.342928 & -0.375 \\
\hline
\end{tabular}
\end{table}

\subsection{Applications to one-dimensional local Hamiltonians}
We benchmark our method on several one-dimensional local Hamiltonians with nearest-neighbor interactions. Specifically, we consider the spin-$1/2$ transverse-field Ising (TFI) model
\begin{equation}
  H_{\text{TFI}}=\sum_i -X_i X_{i+1} - hZ_i,
\end{equation}
the XX model
\begin{equation}
  H_{\text{XX}}=\sum_i X_i X_{i+1} + Y_i Y_{i+1},
\end{equation}
and the XXZ model
\begin{equation}
  H_{\text{XXZ}}=\sum_i X_i X_{i+1} + Y_i Y_{i+1} + \Delta Z_i Z_{i+1}.
\end{equation}
Here $X,Y,Z$ denote the Pauli operators acting on a spin-$1/2$ degree of freedom. These models are useful benchmarks because their thermodynamic-limit ground-state energy densities are known exactly~\cite{PFEUTY197079, takahashi1999thermodynamics}. For the parameter choices considered below, the exact thermodynamic-limit energy density is $-4/\pi$ for the TFI model at $h=1$, $-4/\pi$ for the XX model, and $1-4\ln 2$ for the XXZ model at $\Delta=1$.

We also consider a special type of spin models that reach the maximum quantum violations of certain TI contextuality witnesses~\cite{yangContextualityInfiniteOnedimensional2022}. One type is called the 232-type model meaning that each site has three observables and the interaction is nearest-neighbor, and one such example is \begin{align}
    H=\sum_{i} \sigma^{(i)}_1\sigma^{(i+1)}_0+\sigma^{(i)}_1\sigma^{(i+1)}_1-\sigma^{(i)}_2\sigma^{(i+1)}_0+\sigma^{(i)}_2\sigma^{(i+1)}_1, 
\end{align}
where \begin{equation}\label{232ob-1}{\footnotesize
\begin{aligned}
    &\sigma_0 =     \begin{pmatrix*}[r]
        1 & 0 & 0 & 0 & 0\\
        0 & 1 & 0 & 0 & 0\\
        0 & 0 & -1 & 0 & 0\\
        0 & 0 & 0 & 1 & 0\\
        0 & 0 & 0 & 0 & 1
    \end{pmatrix*},   \sigma_1 = -     \begin{pmatrix*}[r]
        0 & 0 & 1 & 0 & 0\\
        0 & 1 & 0 & 0 & 0\\
        1 & 0 & 0 & 0 & 0\\
        0 & 0 & 0 & 0 & 1\\
        0 & 0 & 0 & 1 & 0
    \end{pmatrix*},
       \sigma_2 =     \begin{pmatrix*}[r]
        1 & 0 & 0 & 0 & 0\\
        0 & 1 & 0 & 0 & 0\\
        0 & 0 & 1 & 0 & 0\\
        0 & 0 & 0 & 1 & 0\\
        0 & 0 & 0 & 0 & -1
    \end{pmatrix*}.
\end{aligned}}
\end{equation}
The exact ground state energy density of the 232-type model has been derived in~\cite{zeng2026}. Unlike the preceding spin-$1/2$ examples, the 232-type model has local dimension $d=5$ and admits an exact matrix-product ground state with bond dimension $4$. It therefore provides a useful benchmark for testing whether the symmetrized MPS method remains effective beyond qubit chains.

\begin{figure}[htbp]
    \centering
    \includegraphics[width=0.9\linewidth]{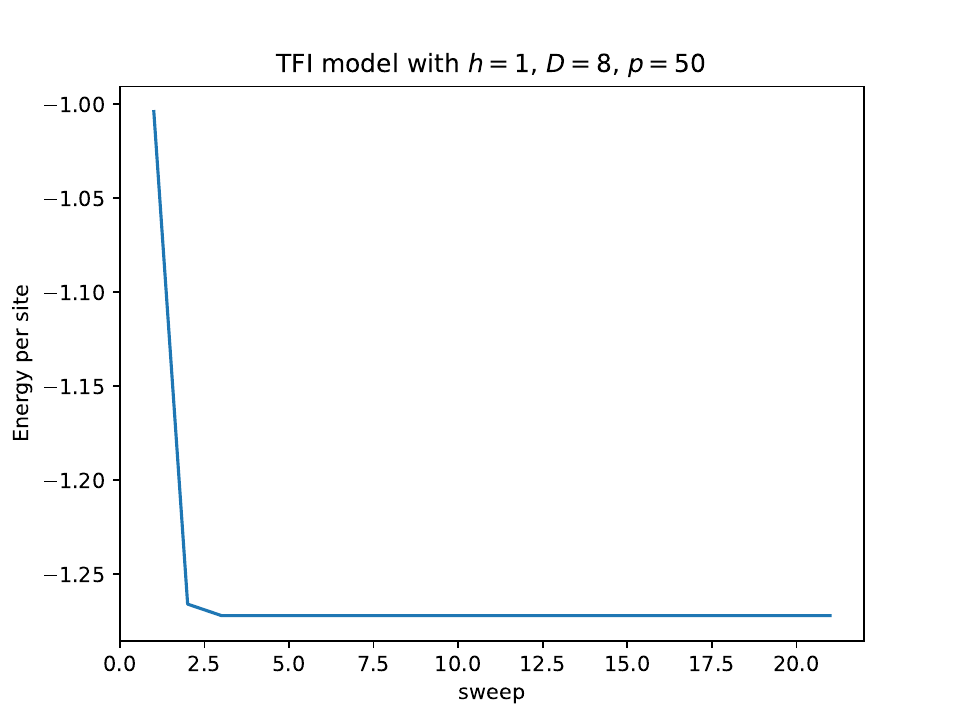}
    \caption{The energy density given by the symmetrized MPS algorithm for the TFI model at $h=1, p=50, D=8$.}
    \label{fig:TFI_1}
\end{figure}

A typical run of the symmetrized MPS algorithm with the TFI model with $h=1$ is shown in Fig.~\ref{fig:TFI_1}. The energy density converges fast to a stable value after a few sweeps. The convergence pattern is similar for the other models, although the number of sweeps required to reach convergence can vary depending on the model and the chosen period $p$.

\begin{figure*}[htbp]
    \centering

    \begin{subfigure}[t]{0.45\textwidth}
        \centering
        \includegraphics[width=\linewidth]{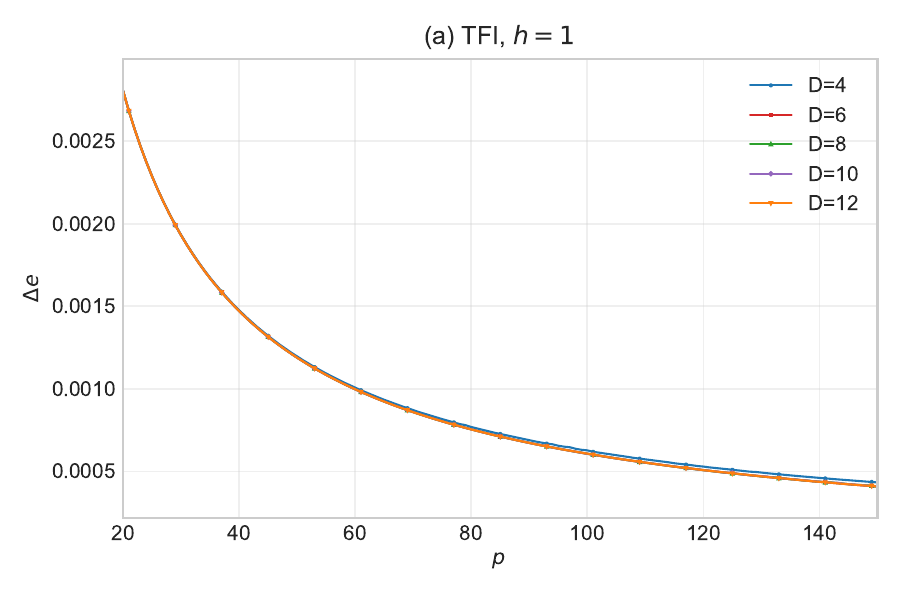}
        \label{fig:heisenberg}
    \end{subfigure}
    \hfill
    \begin{subfigure}[t]{0.45\textwidth}
        \centering
        \includegraphics[width=\linewidth]{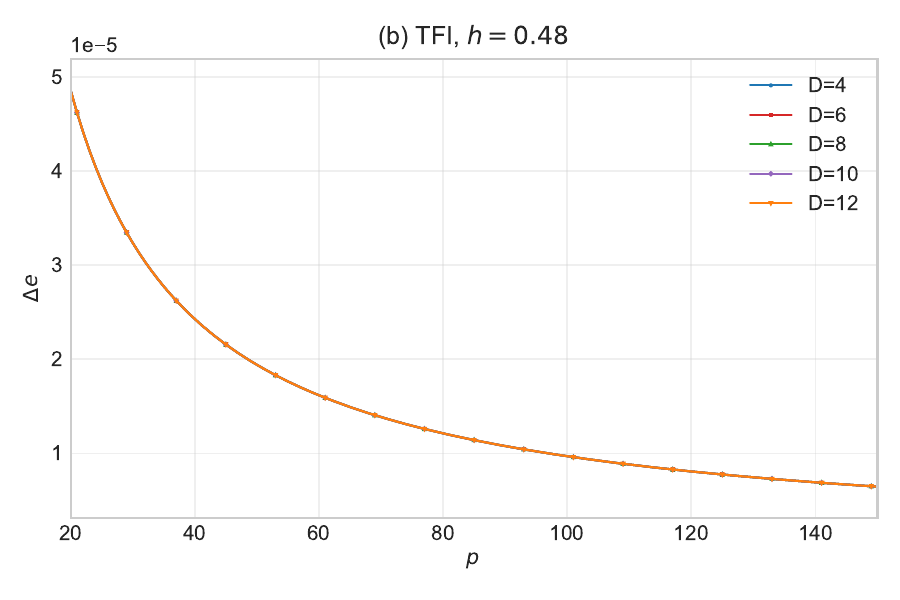}
        \label{fig:ising}
    \end{subfigure}


    \begin{subfigure}[t]{0.45\textwidth}
        \centering
        \includegraphics[width=\linewidth]{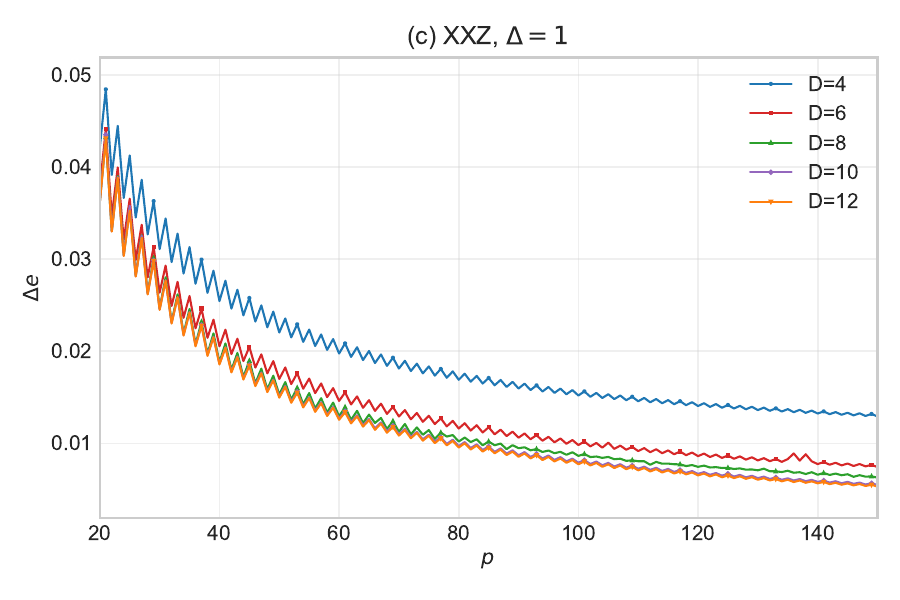}
        \label{fig:XXZ-1}
    \end{subfigure}
    \hfill
    \begin{subfigure}[t]{0.45\textwidth}
        \centering
        \includegraphics[width=\linewidth]{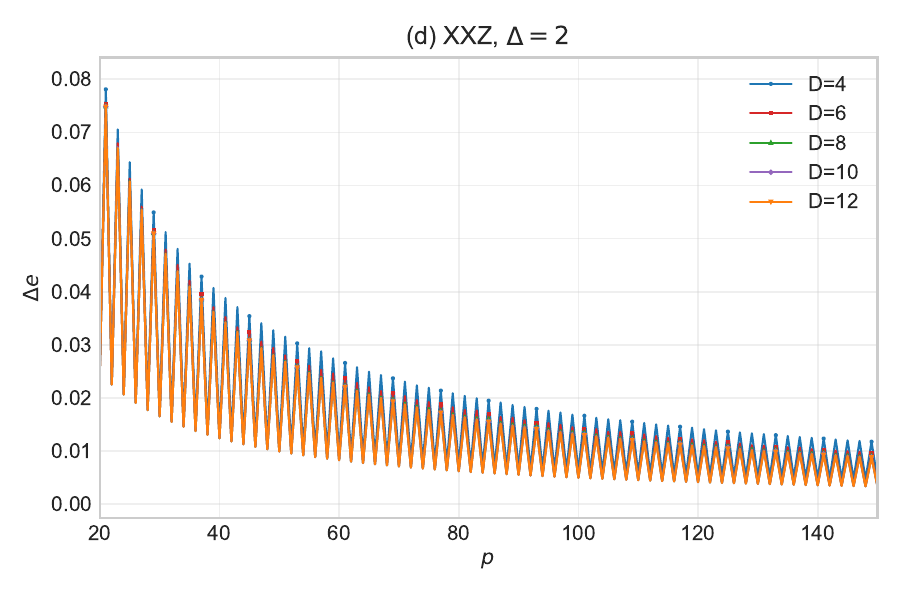}
        \label{fig:XXZ_2}
    \end{subfigure}

    \begin{subfigure}[t]{0.45\textwidth}
        \centering
        \includegraphics[width=\linewidth]{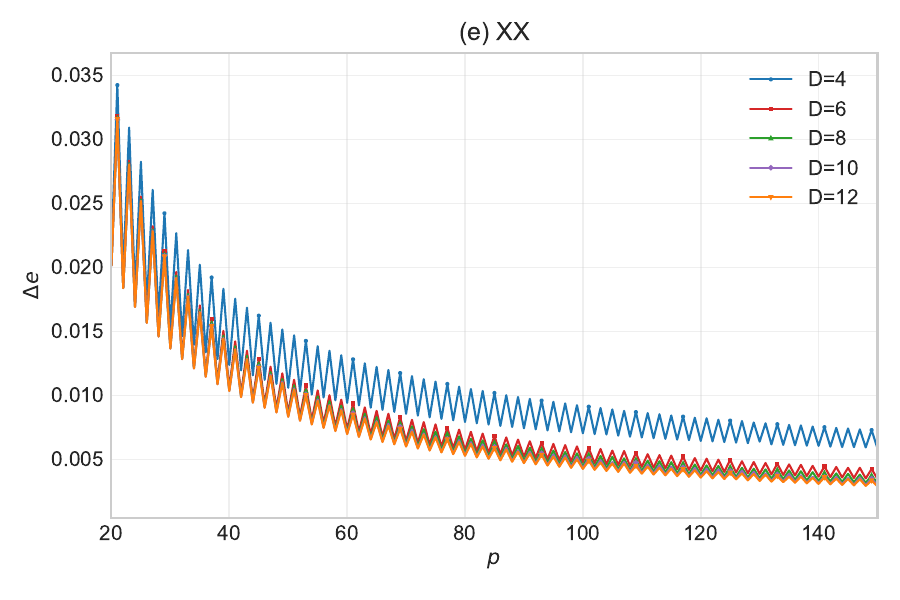}
        \label{fig:XX}
    \end{subfigure}
    \hfill
    \begin{subfigure}[t]{0.45\textwidth}
        \centering
        \includegraphics[width=\linewidth]{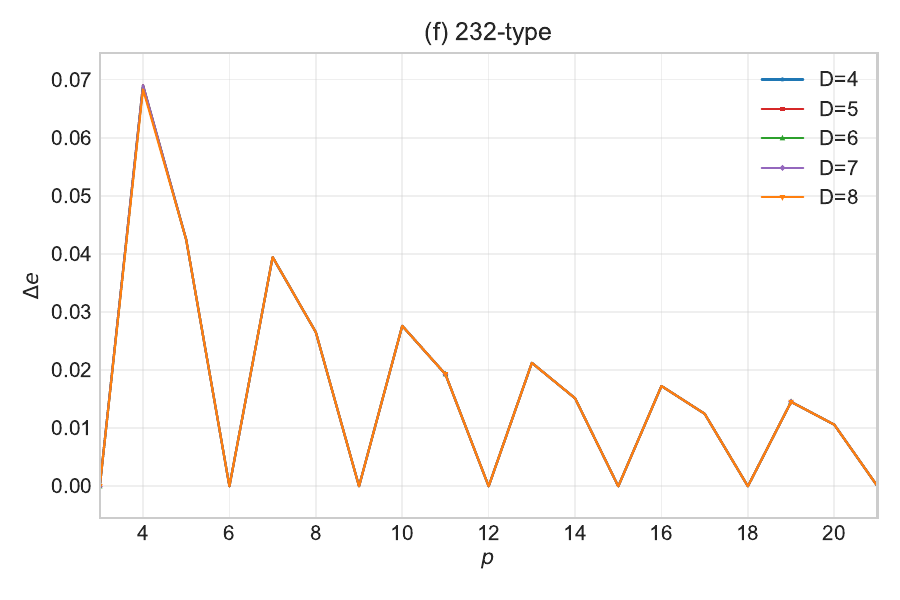}
        \label{fig:232}
    \end{subfigure}

    \caption{The energy error $\Delta e$ given by the symmetrized MPS algorithm. (a) TFI model at $h=1$, (b) TFI model at $h=0.48$, (c) XXZ model at $\Delta=1$, (d) XXZ model at $\Delta=2$, (e) XX model, (f) the 232-type model.}
    \label{fig:four_panels}
\end{figure*}

We perform convergence benchmarks for the models using simple implementations of the symmetrized MPS algorithm. Let $e_p$ denote the variational ground-state energy density obtained with period $p$, and let $e_{\mathrm{exact}}$ denote the exact thermodynamic-limit ground-state energy density. We quantify the error by
$$
\Delta e = e_p - e_{\mathrm{exact}},
$$
and study how this error changes as the period $p$ increases. The results are shown in Fig.~\ref{fig:four_panels}.

Figure~\ref{fig:four_panels} shows that the convergence of the variational energy density depends on both the bond dimension and the period $p$. For small bond dimensions, the symmetrized MPS optimization is more susceptible to local minima. We therefore repeat the calculation with multiple random initializations and report the lowest energy obtained. As the bond dimension increases, this local-minimum issue is no longer observed. In all simulations, convergence is declared when the change in energy between consecutive sweeps is smaller than $\epsilon=10^{-7}$. For $D=4$ and $D=6$, we perform 10 independent runs for each value of $p$ and keep the best result.

The convergence pattern is model dependent. For the TFI model, the energy density approaches the exact value smoothly. By contrast, the XX and XXZ models display oscillatory convergence as a function of $p$, indicating a stronger sensitivity to the chosen period. The difference in convergence behavior can be understood from the dominant correlation structure of the models. The XX and XXZ chains exhibit staggered antiferromagnetic correlations, so finite-period approximations are sensitive to commensurability between the trial period $p$ and the natural correlation wavelength. This produces the observed oscillatory dependence on $p$. In contrast, the transverse-field Ising model considered here is dominated by ferromagnetic correlations, and therefore shows a smoother finite-period convergence. The 232-type model shows a behavior analogous to the MG model ($\Delta e$ is the same for $D=4,5,6,7,8$): the exact value is recovered whenever $p$ is a multiple of $3$, consistent with a period-$3$ ground-state structure.

\subsection{Applications to TI contextuality witnesses}
In this subsection, we apply our SDP-based method to TI contextuality witnesses. We focus on a family of witnesses involving only one- and two-body correlators: the 322-type witnesses, introduced in Ref.~\cite{wangEntanglementNonlocalityInfinite2017, yangContextualityInfiniteOnedimensional2022}. Here the notation $mXY$ means that each party has $X$ measurement settings, each measurement has $Y$ possible outcomes, and the maximum interaction range is $m$.

A 322-type TI contextuality witness has the form
\begin{equation}
\langle\mathcal{E}\rangle_{\text{TI}}=\Bigl\langle\sum_{x=0}^{1}J_x\sigma^{(1)}_x
+\sum_{x,y=0}^1J_{xy}\sigma^{(1)}_x\sigma^{(2)}_y +\sum_{x,z=0}^1J_{xz}\sigma^{(1)}_x\sigma^{(3)}_z\Bigl\rangle_{\text{TI}}
   \geq \mathcal{L}.
\end{equation}
Here the observables $\sigma_x$, $\sigma_y$ and $\sigma_z$ are arbitrary Hermitian operators with eigenvalues $\pm1$, and $\langle\mathcal{E}\rangle_{\text{TI}}$ denotes the expectation value evaluated on TI reduced states. If one can find TI reduced states and observables such that $\langle\mathcal{E}\rangle_{\text{TI}}<\mathcal{L}$, then the witness is violated. In that case, the corresponding quantum system is contextual, in the sense that its statistics cannot be reproduced by any local hidden-variable (LHV) model \cite{bellEinsteinPodolskyRosen1964,nicolasbrunnerBellNonlocality2014,kochenProblemHiddenVariables1967,cabelloBellNonlocalityKochenSpecker2021,budroniKochenSpeckerContextuality2022}.

Finding a violation of a TI contextuality witness requires optimizing over both TI states and local observables in order to obtain a value below the classical bound. To certify that the maximal quantum violation has been reached, one also needs a matching lower bound on the quantum value. Such lower bounds are commonly obtained using the Navascués--Pironio--Acín (NPA) hierarchy \cite{navascuesBoundingSetQuantum2007, navascuesConvergentHierarchySemidefinite2008}. When this lower bound matches the value achieved by an explicit state and a set of observables, the quantum bound is certified.

Consider an $m$-partite inequality $E=\langle\mathcal{E}\rangle\geq\mathcal{L}$. We now recall how the NPA hierarchy gives lower bounds on $E$. In the corresponding Bell scenario, there are $m$ parties, each with $X$ measurement inputs and two outcomes $\pm1$. Let $\sigma_{a_i}^{(i)}$ be the operator acting on site $i$, with $a_i\in\{0,1,\dots,X-1\}$. These operators obey $(\sigma_{a_i}^{(i)})^2=1$ and commute on different sites. The quantity $E$ is the expectation value of the noncommutative polynomial $\mathcal{E}$ specified by the inequality.

Let $(\mathcal{Q}_i)_i$, with $i=1,2,\dots,t$, be a generating sequence of monomials
$$
\mathcal{Q}_i
=
\sigma_{a_1}^{(i_1)}
\sigma_{a_2}^{(i_2)}
\cdots
\sigma_{a_k}^{(i_k)} .
$$
The level of the NPA hierarchy is determined by the maximum degree of the monomials included in this sequence. We choose $(\mathcal{Q}_i)_i$ such that the polynomial $\mathcal{E}$ lies in the linear span of these monomials.

Define the matrix $\mathcal{O}$ by its entries $\mathcal{O}_{i,j}=\mathcal{Q}_i^\dagger\mathcal{Q}_j$. Let $(M_k)_k$, with $k=1,2,\dots,s$, be a Hermitian basis for the vector space spanned by the entries of $\mathcal{O}$, chosen such that $M_1=1$. Then $\mathcal{E}$ and $\mathcal{O}$ can be decomposed as
\begin{equation}\label{eq:npa-basis}
\mathcal{E} = \sum_{k=1}^s c_k M_k,\quad \mathcal{O} = \sum_{k=1}^s A_k M_k,
\end{equation}
where each $A_k$ is a $t\times t$ Hermitian matrix. Writing $x_k=\langle M_k\rangle$, the moment matrix and the objective value are
\begin{equation}\label{eq:npa-gamma}
\Gamma = \sum_{k=1}^s A_k x_k, \quad E = \sum_{k=1}^s c_k x_k.
\end{equation}

By construction, the moment matrix $\Gamma$ is positive semidefinite. This gives an SDP relaxation of the quantum bound:
\begin{equation} \label{npa-org}
    \begin{aligned} 
    \min_{x_2,\ldots ,x_s} \quad & \sum_{k=1}^s c_k x_k \\
    \text{subject to} \quad & x_1 = 1, \quad
\Gamma = \sum_{k=1}^s A_k x_k \succeq 0.
\end{aligned}
\end{equation}
Eq.~\eqref{npa-org} provides a lower bound on the quantum value, and the bound becomes tighter as larger sets of monomials are included. Although one generally cannot know in advance which level is sufficient, low levels of the hierarchy often already reproduce the quantum limit in practice. 

For TI contextuality witnesses, however, an additional constraint must be addressed: the admissible reduced states must come from infinite TI states. Since the set of such TI reduced states is not semialgebraic \cite{blakajSetReducedStates2024}, it cannot be characterized exactly by SDP constraints. One possible approach is to impose local translation invariance (LTI) within the NPA hierarchy, which gives an outer approximation to the TI quantum bound, as used in Ref.~\cite{yangContextualityInfiniteOnedimensional2022}.

Here we take a complementary approach: instead of relaxing the TI constraint by local translation invariance, we impose a finite-period block-product structure within the NPA hierarchy. We refer to the resulting construction as the periodic-NPA hierarchy. At a finite relaxation level, the resulting SDP
should be interpreted as a relaxation of the corresponding periodic variational problem; by itself, its bound direction relative to the TI quantum minimum is not automatic. Tightness can be established when the SDP value is
matched by an explicit periodic physical realization.

To incorporate $p$-periodic block-product states into the NPA hierarchy, we consider a finite system of $kp$ parties, where $k\in\mathbb{N}$, and group every $p$ consecutive parties into one block. The block-product periodic ansatz requires these $k$ blocks to represent identical copies of a single $p$-site unit cell, with different blocks forming a block-product structure. Following the moment-matrix construction of Ref.~\cite{moroderDeviceindependentEntanglementQuantification2013}, we encode this structure at the level of the NPA moment matrix. Let $\Lambda=\sum_{s\in \Omega}\ket{s}\bra{0}\otimes s$, where $\Omega$ contains the identity and the chosen single-site operators. The moment matrix $\Gamma$ is then defined as
\begin{widetext}
\begin{equation}\label{gamma-dps}
\begin{aligned}
    \Gamma&=\tr_{\rho}(\Lambda^{\otimes pk}(\ket{0}\bra{0})^{\otimes pk}\otimes \rho \Lambda^{\dagger\otimes pk})\\
    &=\sum_{\substack{s_1,s_2,\cdots, s_{kp}\in \Omega,\\s_1',s_2',\cdots, s_{kp}'\in \Omega}}\ket{s_1s_2\cdots s_{kp}}\bra{s_1's_2'\cdots s_{kp}'}\tr(s_1\otimes \cdots \otimes s_{kp}\rho s_1'\otimes \cdots \otimes s_{kp}')\\
    &=\sum_{\substack{s_1,s_2,\cdots, s_{kp}\in \Omega,\\s_1',s_2',\cdots, s_{kp}'\in \Omega}}\ket{s_1s_2\cdots s_{kp}}\bra{s_1's_2'\cdots s_{kp}'}\langle s_1s_1^{\dagger}\cdots s_{kp}s_{kp}^{\dagger}\rangle.
\end{aligned}
\end{equation}
\end{widetext}

The moment matrix $\Gamma$ must satisfy two additional constraints that encode the periodic block structure. First, it should be positive under partial transposition with respect to each block. This condition is a necessary relaxation of the block-product structure. Second, it should be invariant under permutations of the $k$ blocks, which enforces that all blocks are treated as identical copies of the same $p$-site unit cell.

The symmetrized objective function is
\begin{equation}
\begin{aligned}
\langle\mathcal{E}_{\text{sym}} \rangle
= \frac{1}{p}\sum_{i=1}^{p} \Biggl\langle
&\sum_{x=0}^{1}J_x\sigma^{(i)}_x
+\sum_{x,y=0}^1J_{xy}\sigma^{(i)}_x\sigma^{(i+1)}_y \\
&+\sum_{x,z=0}^1J_{xz}\sigma^{(i)}_x\sigma^{(i+2)}_z
\Biggr\rangle .
\end{aligned}
\end{equation}
Since $\langle\mathcal{E}_{\text{sym}}\rangle$ is linear in the entries of $\Gamma$, there exists a coefficient matrix $C$ such that
$\langle\mathcal{E}_{\text{sym}}\rangle=\tr(C\Gamma)$. The generating sequence of $\Gamma$ is $\{\bigotimes_{1}^{kp} s_i, s_i\in \Omega\}$. By choosing a Hermitian basis, we can find matrices $\{A_k\}$ such that $\Gamma = \sum_{k=1}^s A_k x_k$ in the same way as Eq.~\eqref{eq:npa-basis} and Eq.~\eqref{eq:npa-gamma}. The periodic-NPA relaxation is therefore the following SDP:
\begin{equation}
	\label{dps-npa}
	\begin{aligned}
		\min_{x_2,\ldots,x_s} \quad&\tr(C\Gamma) \\
		\text{s.t.}\quad & \Gamma = \sum_{k=1}^s A_k x_k \succeq 0, x_1=1,\\
        & \Gamma^{T_S} \succeq 0, \quad \forall \emptyset \neq S \subsetneq \{1,\dots,k\}, \\\\
        & \pi(\Gamma)=\Gamma, \forall \pi\in S_k,
	\end{aligned}
\end{equation}
where $\Gamma^{T_S}$ denotes the partial transpose of $\Gamma$ with respect to the subset $S$, $\pi$ denotes a permutation of the $k$ blocks, and $S_k$ is the symmetric group acting on these blocks.


In our implementation, we set \(k=2\). For computational efficiency, we do
not construct the full \(3^{2p}\times 3^{2p}\) moment matrix. Instead, we use
a truncated operator set. We first include all operators appearing in
\(\mathcal{E}_{\mathrm{sym}}\), and then enlarge this set by closing it under
the partial-transposition and block-permutation operations entering the
periodic-NPA constraints. This ensures that the truncated moment matrix is
compatible with the imposed periodic-NPA symmetries.

We apply this construction to all inequalities listed in
Table~\ref{tab:322}, taken from Ref.~\cite{yangContextualityInfiniteOnedimensional2022}. For the contextuality witness with $p=4$, the moment matrix has size $841\times841$, while for those with $p=5$, it has size $2209\times2209$. The periodic-NPA SDPs were solved using MOSEK \cite{mosek}, with the solver tolerance set to $10^{-8}$. In all reported computations, MOSEK returned optimal primal- and dual feasible solutions. The solver-reported primal and dual feasibility residuals were typically of order $10^{-12}$ and $10^{-9}$, respectively, with negligible primal-dual objective gaps.

For all ten inequalities, the periodic-NPA optima agree numerically with the corresponding quantum limits reported in Ref.~\cite{yangContextualityInfiniteOnedimensional2022}. Since a finite-level periodic-NPA problem is still a relaxation of the exact periodic variational problem, this agreement alone does not certify that the relaxation value is physically attainable. We therefore also construct a physical moment matrix \(\Gamma_{\mathrm{phys}}\) using the optimal observables and ground states from Ref.~\cite{yangContextualityInfiniteOnedimensional2022}. In all ten cases, the physical moment matrices agree with the SDP optimizers up to Frobenius-norm discrepancies below $10^{-8}$. This matching provides numerical evidence that the finite-level periodic-NPA relaxation is tight in these examples, and indicates that the optimal quantum violations can be realized by block-product periodic states.

\begin{table*}[htbp]
    \caption{The periods of the 10 322-type inequalities that are identified with maximum quantum violation. The $J$ coefficients are the coefficients of the inequalities. $\mathcal{L}$ is the classical bound. $\mathcal{Q}$ is the quantum limit certified in Ref.~\cite{yangContextualityInfiniteOnedimensional2022}. $p$ is the period, and $Q_p$ is the periodic-NPA value.}
    \centering
    \renewcommand\arraystretch{1.1}
    \setlength{\tabcolsep}{1.1mm}{
    \begin{tabular}{|c|c|c|c|c|c|c|c|c|c|c|c|c|c|c|c|c|c|c|c|c|}	
    \hline
No. & $J_0$ &$J_1$ & $J_{00}^{12}$ & $J_{01}^{12}$ & $J_{10}^{12}$ & $J_{11}^{12}$ & $J_{00}^{13}$ & $J_{01}^{13}$ & $J_{10}^{13}$ & $J_{11}^{13}$& $\cal{L}$ & ${\cal{Q}}$ & $p$ & $Q_p$
\\ \hline						
    1 &	 -6&  0&  2 &	3 &	3 &	-2 &	3 &	-1 &	-1 &	1 & -6 & -6.32747 & 4	& -6.32747
    \\ \hline									 	 	 
    2&	-4 &	2 &	2 &	2 &	2 &	-4 &	1 &	-1 &	-1 &	3 & -6 & -6.33711 & 4 & -6.33712
    \\	\hline								 	 	
    3&	-3 &	1 &	1 &	1 &	1 &	-1 &	1 & 0 &	-1 & 1 &  -3	& -3.20710 & 4 & -3.20711
    \\ \hline
    4&	-2 &	-2 &	-2 &	1 &	-1 &	-2 &	1 &	0 &	2 &	1 &  -4	& -4.14623 & 4 & -4.14623
    \\ \hline				 	
    5	& 	-11 &	1 &	5 &	2 &	2 &	-1& 	4 &	-1 &	-2 	&1  &-8 &	-8.12123& 4 & -8.12123
    \\ \hline									 		
    6&	-3 &	-3 &	2 &	2 	&-1 &	2 &	1 &	1 &	-1 &	0 & -4 & -4.10309&5		& -4.10310
\\ \hline
7&		-3 &	-3& 	2 &	2 &	2 &	-3 &	1 &	0 &	-1 &	2 & -5	&-5.29851& 5	&-5.29852
\\ \hline
8&	-2&	 	-4 &		-2 &		2 &		2 &		2 	&	1 &		0 &		0 &		1 &-4& -4.33137& 5	& -4.33137
\\ \hline
9	&			0 &		-4 &		2 &		2 &		-2 &		2 &		0 &		1 &		-1 &		0 	& -4	& -4.41421& 4 & -4.41421	 \\ \hline	
10	&			-2 &		2 &		2 &		-2 &		-2 &		-4 &		1 &		1 &		1 &		2 & -5	& -5.26969& 5 & -5.26969\\ \hline
    \end{tabular}}
    \label{tab:322}
\end{table*}

\section{Conclusion}\label{sec:conclusion}

We introduced the block-product periodic marginal problem, which asks whether a local reduced density matrix can be obtained from the translation average of an infinite block-product periodic state. This formulation connects local compatibility, hidden translation-symmetry breaking, and variational estimates of ground-state energy densities in infinite TI systems.

We developed two approaches to this problem. The SDP-based method provides a certified relaxation for small systems, and correctly recovers the $2$-periodic structure and exact energy density of the Majumdar--Ghosh model for compatible periods. The symmetrized MPS method gives a scalable variational ansatz for larger periods. Benchmarks on several one-dimensional Hamiltonians show that the variational energy improves with increasing period and bond dimension, while the convergence pattern depends on the model.

We also applied the periodic-extension viewpoint to TI contextuality witnesses by constructing a periodic version of the NPA hierarchy. For the tested inequalities, the periodic-NPA values reproduce the known quantum limits, indicating that the corresponding optimal violations can be realized by block-product periodic states.

These results suggest that periodic extensions provide a useful intermediate framework between fully TI descriptions and symmetry-broken many-body structures. Future work includes improving the convergence and stability of the symmetrized MPS algorithm, establishing sharper convergence guarantees for the SDP hierarchy, and extending the approach to higher-dimensional tensor-network ansätze.

\section*{Acknowledgments}
The authors acknowledge Miguel Navascu\'{e}s and Paulo Abiuso for discussions. This work was supported by the NSFC (No.~12574536) and the Sichuan Science and Technology Program (2024YFHZ0371).

\bibliographystyle{unsrt}
\bibliography{ref}
\end{document}